\documentclass[11pt]{article}
\usepackage{amssymb,amsmath,amsfonts}
\usepackage{graphicx}
\usepackage{graphics}
\usepackage{eepic,epsfig}

\makeatletter
\@addtoreset{equation}{section}

\makeatother

\begin{document}

\title{Fermionic vacuum currents in topologically nontrivial braneworlds:
Two-brane geometry}
\author{ S. Bellucci$^{1}$\thanks{%
E-mail: bellucci@lnf.infn.it }, A. A. Saharian$^{2}$\thanks{%
E-mail: saharian@ysu.am }, H. G. Sargsyan$^{2}$\thanks{%
E-mail: hayk28.sargsyan@yahoo.com}, V. V. Vardanyan$^{2,3}$\thanks{%
E-mail: vardanyan@lorentz.leidenuniv.nl } \vspace{0.3cm} \\
\textit{$^1$ INFN, Laboratori Nazionali di Frascati,}\\
\textit{Via Enrico Fermi 40, 00044 Frascati, Italy} \vspace{0.3cm}\\
\textit{$^2$ Department of Physics, Yerevan State University,}\\
\textit{1 Alex Manoogian Street, 0025 Yerevan, Armenia }\vspace{0.3cm}\\
\textit{$^3$ Lorentz Institute for Theoretical Physics, Leiden University,}\\
\textit{2333 CA Leiden, Netherlands}}
\maketitle

\begin{abstract}
The vacuum expectation value (VEV) of the fermionic current density is
investigated in the geometry of two parallel branes in locally AdS spacetime
with a part of spatial dimensions compactified to a torus. Along the toral
dimensions quasiperiodicity conditions are imposed with general phases and
the presence of a constant gauge field is assumed. The influence of the
latter on the VEV is of the Aharonov-Bohm type. Different types of boundary
conditions are discussed on the branes, including the bag boundary condition
and the conditions arising in $Z_{2}$-symmetric braneworld models. Nonzero
vacuum currents appear along the compact dimensions only. In the region
between the branes they are decomposed into the brane-free and brane-induced
contributions. Both these contributions are periodic functions of the
magnetic flux enclosed by compact dimensions with the period equal to the
flux quantum. Depending on the boundary conditions, the presence of the
branes can either increase or decrease the vacuum current density. For a
part of boundary conditions, a memory effect is present in the limit when
one of the branes tends to the AdS boundary. Unlike to the fermion
condensate and the VEV of the energy-momentum tensor, the VEV of the current
density is finite on the branes. Applications are given to
higher-dimensional generalizations of the Randall-Sundrum models with two
branes and with toroidally compact subspace. The features of the fermionic
current are discussed in odd-dimensional parity and time-reversal symmetric
models. The corresponding results for three-dimensional spacetime are
applied to finite length curved graphene tubes threaded by a magnetic flux.
It is shown that a nonzero current density can also appear in the absence of
the magnetic flux if the fields corresponding to two different points of the
Brillouin zone obey different boundary conditions on the tube edges.
\end{abstract}

PACS numbers: 04.62.+v, 03.70.+k, 98.80.-k, 61.46.Fg

\bigskip

\section{Introduction}

\label{sec:introd}

In quantum field theory the vacuum is defined as a state of quantum fields
with the zero number of quanta. It depends on the choice of the complete set
of mode functions in terms of which the quantization of fields is done (see,
for instance, \cite{Birr82}). The mode functions and, as a consequence of
that, the properties of the vacuum are sensitive to both the local and
global characteristics of the background geometry. In particular, the vacuum
expectation values (VEVs) of physical observables depend on the boundary
conditions imposed on fields. This dependence is manifested in the Casimir
effect \cite{Most97} and has been investigated for large number of bulk and
boundary geometries. The boundary conditions may have different physical
origins. They can be induced by the presence of boundaries (material
boundaries in quantum electrodynamics, domain walls separating different
phases, horizons in gravitational physics, branes in braneworld scenarios)
or as a consequence of nontrivial spatial topology. In some models
formulated in background of manifolds with edges, the imposition of boundary
conditions on those edges is required to prevent the unitarity of the
theory. In the present paper we consider a physical problem with different
sources for the polarization of vacuum. They include the background
gravitational field, gauge field, boundaries and nontrivial spatial topology.

The background geometry we are going to discuss is locally anti-de Sitter
(AdS) one. Being the maximally symmetric solution of the vacuum Einstein
equations with a negative cosmological constant, AdS spacetime is among the
most popular geometries in quantum field theory on curved backgrounds. First
of all, because of high symmetry, a large number of physical problems are
exactly solvable on that background. These solutions may help to shed light
on the influence of gravitational field on quantum matter in less symmetric
geometries. The lack of global hyperbolicity and the presence of the modes
regular and irregular on the AdS boundary give rise to new principal
questions in the quantization procedure of fields having no analogs in
quantum field theory on the Minkowski bulk. The importance of the
corresponding investigations is also related to the fact that the AdS
spacetime naturally appears as a ground state in extended supergravity and
in string theories and also as the near horizon geometry of the extremal
black holes and domain walls.

The further increase of the interest to the AdS based field theories is
motivated by a crucial role of the corresponding geometry in two exciting
developments of theoretical physics in the past decade. The first one is the
braneworld scenario with large extra dimensions which provides a geometrical
solution to the hierarchy problem between the gravitational and electroweak
energy scales \cite{Maar10}. The corresponding models are usually formulated
on higher-dimensional AdS bulk with branes parallel to the AdS boundary and
the weak coupling of 4-dimensional gravity is generated by the large
physical volume of extra dimensions. Braneworlds naturally appear in the
string/M theory context and present a novel setting for the discussion of
phenomenological and cosmological issues related to extra dimensions. The
second development is related to the AdS/CFT correspondence (for reviews see
\cite{Ahar00}) that relates string theories or supergravity in the AdS bulk
with a conformal field theory living on its boundary. This duality between
two different theories has many interesting consequences and provides a
powerful tool for the investigation of gauge field theories. Among the
recent developments of the AdS/CFT correspondence is the application to
strong-coupling problems in condensed matter physics (familiar examples
include holographic superconductors, quantum phase transitions and
topological insulators) \cite{Pire14}.

In the present paper the global properties of the background geometry will
be different from those for AdS spacetime. It will be assumed that a part of
the Poincar\'{e} coordinates in the AdS line element are compactified on a
torus. In addition, we assume the presence of two branes parallel to the AdS
boundary. As a consequence, two types of conditions are imposed on the
operators of quantum fields: periodicity conditions along compact dimensions
and boundary conditions on the branes. In the Randall-Sundrum type
branewrolds the latter are dictated by the $Z_{2}$-symmetry with respect to
the branes. Both these conditions modify the spectrum of vacuum fluctuations
and give rise to the Casimir type contributions in the physical
characteristics of the vacuum state. In particular, motivated by the problem
of radion stabilization in braneworld scenario, the brane-induced quantum
effects have been intensively investigated for scalar \cite{Fabi00},
fermionic \cite{Flac01}-\cite{Eliz13} and vector fields \cite{Teo10}. The
models with de Sitter branes have been considered in \cite{Noji00}. The
Casimir effect in AdS spacetime with additional compact subspaces is
discussed in \cite{Flac03}. The expectation value of the surface
energy-momentum tensor for a scalar field, induced on branes, and related
cosmological constant are studied in \cite{Saha04}.

The papers cited above consider mainly the vacuum energy or the VEV\ of the
energy-momentum tensor. For charged fields, an important local
characteristics are the expectation values of the charge and current
densities. For scalar and fermionic fields in flat backgrounds with a part
of spatial dimensions compactified to a torus, these expectation values at
zero and finite temperatures were considered in Refs. \cite%
{Beze13c,Bell10,Bell14}. The results for fermionic fields in a special case
of two spatial dimensions have been applied to cylindrical and toroidal
carbon nanotubes described in terms of the long-wavelength effective Dirac
model. The boundary-induced effects of the Casimir type on the vacuum
charges and currents are discussed in \cite{Bell13,Bell15}. The fermionic
current density induced by a magnetic flux in planar rings with concentric
circular boundaries has been investigated in \cite{Rech07}. The persistent
currents in normal metal rings having a similar physical origin have been
experimentally observed in \cite{Bluh09}. The effects of edges on the
fermion condensate and the currents in two-dimensional conical spaces are
discussed in \cite{Beze10C}. More complicated problems for the vacuum
currents in locally de Sitter and AdS background geometries with toroidally
compactified spatial dimensions are considered in \cite{Bell13b} and \cite%
{Beze15,Bell17}. Induced current in AdS spacetime in the presence of a
cosmic string and compactified spatial dimension is studied in \cite{Oliv18}%
. The brane-induced effects on the the current density for a charged scalar
field with Robin boundary conditions in locally AdS bulk are investigated in
\cite{Bell15b,Bell16}. The corresponding problem for a fermionic field in
the geometry of a single brane with bag boundary condition has been
considered in \cite{Bell18}. Continuing in this line of investigations, here
we consider the fermionic vacuum currents for two-brane geometry in
background of locally AdS spacetime with compact dimensions and for
different combinations of the boundary conditions on them.

The organization of the paper is as follows. In the next section we specify
the bulk and boundary geometries, the topology and the boundary conditions
imposed on the field. In Section \ref{sec:Modes}, a complete set of the
positive and negative energy solutions to the Dirac equation is presented in
the region between two branes and the eigenvalues of the radial quantum
number are specified. The VEV of the current density for the bag boundary
condition on the branes is investigated in Section \ref{sec:Curr}. Two
alternative representations are provided and the asymptotic behavior is
discussed in various limiting regions of the parameters. In Section \ref%
{sec:jlBC2} we consider the VEV of the current density for another type of
boundary condition that differs from the bag boundary condition by the sign
of the term containing the normal to the boundary. In Section \ref{sec:Brane}%
, the fermionic current in $Z_{2}$-symmetric braneworld models with two
branes is investigated. Various combinations of the boundary conditions,
dictated by the $Z_{2}$-symmetry with respect to the branes, are discussed.
The features of the fermionic current in odd-dimensional parity and
time-reversal symmetric models are considered in Section \ref{sec:OddD} and
applications are given to the Dirac model describing the long wavelength
properties of curved graphene tubes. The main results are summarized in
Section \ref{sec:Conc}. In Appendix \ref{sec:App}, by using the generalized
Abel-Plana formula, a summation formula is derived for series over the
eigenmodes of the fermionic field in the region between the branes for
boundary conditions on the field operator discussed in the main text.

\section{Problem setup}

\label{sec:Geom}

In this section we describe the bulk and boundary geometries, the field and
the periodicity and boundary conditions.

\subsection{Background geometry}

Consider a $(D+1)$-dimensional spacetime with the line element
\begin{equation}
ds^{2}=e^{-2y/a}\eta _{ik}dx^{i}dx^{k}-dy^{2}=(a/z)^{2}\eta _{\mu \nu
}dx^{\mu }dx^{\nu },  \label{ds2}
\end{equation}%
where $a$ is a constant having the dimension of length, the Latin and Greek
indices run over $0,1,\ldots ,D-1$ and $0,1,\ldots ,D$, respectively, $\eta
_{\mu \nu }=\mathrm{diag}(1,-1,\ldots ,-1)$ is the metric tensor for the
Minkowski spacetime in the Cartesian coordinates. The conformal coordinate $%
z=x^{D}$ is expressed in terms of the coordinate $y$, $-\infty <y<+\infty $,
by the relation $z=ae^{y/a}$ with the range $0\leqslant z<\infty $. The line
element (\ref{ds2}) coincides with the one for the $(D+1)$-dimensional AdS
spacetime, described in Poincar\'{e} coordinates. In the case of AdS
spacetime, for the coordinates $x^{i}$, $i=1,\ldots ,D$, one has $-\infty
<x^{i}<+\infty $. The global properties of the geometry we are going to
consider here will be different. Namely, we assume that the subspace with
the coordinates $\mathbf{x}_{(q)}=(x^{p+1},\ldots ,x^{D-1})$, $q=D-p-1$, is
compactified to a $q$-dimensional torus $T^{q}=(S^{1})^{q}$ with the lengths
of the compact dimensions $L_{l}$, $0\leqslant x^{l}\leqslant L_{l}$, $%
l=p+1,\ldots ,D-1$. In what follows we will denote by $V_{q}=L_{p+1}\cdots
L_{D-1}$ the volume of the compact subspace. For the coordinates $\mathbf{x}%
_{(p)}=(x^{1},\ldots ,x^{p})$, as usual, one has $-\infty <x^{i}<+\infty $, $%
i=1,\ldots ,p$, and, hence, the subspace covered by the set of coordinates $(%
\mathbf{x}_{(p)},\mathbf{x}_{(q)})=(x^{1},\ldots ,x^{D-1})$ has topology $%
R^{p}\times T^{q}$. Note that the constant $L_{l}$ is the coordinate length
of the $l$th compact dimension. The physical (or proper) length $L_{(p)l}$
of that dimension, measured by an observer having a fixed $z$ coordinate, is
given by $L_{(p)l}=(a/z)L_{l}$ and it decreases with increasing $z$ (see
figure \ref{fig5} below for the $D=2$ spatial geometry embedded in a
three-dimensional Euclidean space).

The last relation in (\ref{ds2}) shows that the geometry under consideration
is conformally related to the half (with $0\leqslant x^{D}<\infty $) of the
locally Minkowskian $(D+1)$-dimensional spacetime with spatial topology $%
R^{p+1}\times T^{q}$. The Minkowskian counterpart contains a boundary $%
x^{D}=0$ the boundary condition on which is determined by the boundary
condition imposed on the AdS boundary $z=0$. The AdS horizon is presented by
the hypersurface $z=\infty $. The toroidal compactification under
consideration does not change the local geometry and the Ricci tensor $%
R_{\mu }^{\nu }=-D\delta _{\mu }^{\nu }/a^{2}$ is the same as that for AdS
spacetime.

As a boundary geometry we will assume the presence of two co-dimension one
branes located at $y=y_{1}$ and $y=y_{2}$, $y_{1}<y_{2}$. For the
corresponding values of the conformal coordinate $z$ one has $%
z_{j}=ae^{y_{j}/a}$, $j=1,2$. Note that the physical distance between the
branes is given by $y_{2}-y_{1}=a\ln (z_{2}/z_{1})$ and they have spatial
topology $R^{p}\times T^{q}$. For the extrinsic curvature tensor of the
brane at $z=z_{j}$ one has $K_{ik}^{(j)}=\pm g_{ik}/a$, where the upper and
lower signs correspond to the regions $z\leqslant z_{j}$ and $z\geqslant
z_{j}$. As a consequence of the nonzero extrinsic curvature, the physical
effects of the brane on the properties of the quantum vacuum are different
in those regions. In the generalized Randall-Sundrum type models with
additional compact dimensions, the hypersurfaces $y=y_{1}$ and $y=y_{2}$
correspond to the hidden and visible branes, respectively. Higher
dimensional generalizations of the braneworld models with compact dimensions
are, in particular, important from the viewpoint of underlying fundamental
theories in higher dimensions such as superstring/M theories. The
consideration of more general spacetimes may provide interesting extensions
of the Randall-Sundrum mechanism for the geometric origin of the hierarchy.

\subsection{Field and boundary conditions}

Having specified the bulk and boundary geometries, now we pass to the field
content. We consider a charged fermionic field $\psi (x)$ with the mass
parameter $m$ in the presence of an external classical abelian gauge field $%
A_{\mu }(x)$. Here and in what follows the shorthand notation $%
x=(x^{0}=t,x^{1},\ldots ,x^{D})$ is used for the spacetime coordinates. The
coupling parameter between the fermionic and gauge fields will be denoted by
$e$. For a fermionic field realizing the irreducible representation of the
Clifford algebra the number of components of the spinor $\psi (x)$ is equal
to $N=2^{[(D+1)/2]}$, where the square brackets mean the integer part.
Introducing the gauge extended covariant derivative operator $D_{\mu
}=\partial _{\mu }+\Gamma _{\mu }+ieA_{\mu }$, with $\Gamma _{\mu }$ being
the spin connection, the field equation is written as
\begin{equation}
\left( i\gamma ^{\mu }D_{\mu }-m\right) \psi (x)=0.  \label{Feq}
\end{equation}%
The curved spacetime $N\times N$ Dirac matrices are expressed in terms of
the corresponding flat spacetime matrices $\gamma ^{(b)}$ by the relation $%
\gamma ^{\mu }=e_{(b)}^{\mu }\gamma ^{(b)}$, where $e_{(b)}^{\mu }$ are the
vierbein fields. In the conformal coordinates $x^{\mu }$, with $x^{D}=z$,
the vierbein fields can be taken in the form $e_{(b)}^{\mu }=(z/a)\delta
_{b}^{\mu }$. With this choice, for the components of the spin connection
one gets $\Gamma _{D}=0$ and $\Gamma _{k}=\eta _{kl}\gamma ^{(D)}\gamma
^{(l)}/(2z)$ for $k=0,\ldots ,D-1$. The field equation (\ref{Feq}) is
invariant under the gauge transformation $\psi (x)=\psi ^{\prime
}(x)e^{-ie\chi (x)}$, $A_{\mu }(x)=A_{\mu }^{\prime }(x)+\partial _{\mu
}\chi (x)$.

The background geometry is not simply connected and, in addition to the
field equation, the periodicity conditions on the field operator should be
specified along compact dimensions for the theory to be defined. For the $l$%
th compact dimension we will impose the condition
\begin{equation}
\psi (t,\mathbf{x}_{(p)},\mathbf{x}_{(q)}+L_{l}\mathbf{e}_{(l)},x^{D})=e^{i%
\alpha _{l}}\psi (t,\mathbf{x}_{(p)},\mathbf{x}_{(q)},x^{D}),  \label{PCo}
\end{equation}%
where $\mathbf{e}_{(l)}$ is the unit vector along the dimension $x^{l}$ with
the components $e_{(l)}^{i}=\delta _{l}^{i}$ and $\alpha _{l}$, $%
l=p+1,\ldots ,D-1$, are constant phases. The special cases, most frequently
considered in the literature, correspond to untwisted ($\alpha _{l}=0$) and
twisted ($\alpha _{l}=\pi $) fields. The periodicity conditions with $\alpha
_{l}\neq 0$ have been used in the literature to exclude the zero mode of the
fermionic field. For the gauge field the simplest configuration will be
assumed with $A_{\mu }=\mathrm{const}$. Only the components $A_{l}$, $%
l=p+1,\ldots ,D-1$, along compact dimensions are physically relevant. Their
effects on physical observables are of the Aharonov-Bohm type and they are
induced by the nontrivial topology of the background geometry. By the gauge
transformation $\{\psi (x),A_{\mu }\}\rightarrow \{\psi ^{\prime }(x),A_{\mu
}^{\prime }\}$, with the transformation function $\chi =b_{\mu }x^{\mu }$
and constant $b_{\mu }$, one gets a new set of fields $\{\psi ^{\prime
}(x),A_{\mu }^{\prime }\}=\{\psi (x)e^{ieb_{\mu }x^{\mu }},A_{\mu }-b_{\mu
}\}$. The periodicity conditions for the field $\psi ^{\prime }(x)$ are of
the form (\ref{PCo}) with new phases $\alpha _{l}^{\prime }=\alpha
_{l}+eb_{l}L_{l}$. Hence, by the gauge transformation the set of parameters $%
\{\alpha _{l},A_{l}\}$ is transformed to a new set $\{\alpha _{l}^{\prime
},A_{l}^{\prime }\}=\{\alpha _{l}+eb_{l}L_{l},A_{l}-b_{l}\}$. In what
follows, it is convenient for us to work in the gauge with $b_{\mu }=A_{\mu
} $ with the zero vector potential $A_{\mu }^{\prime }$. The corresponding
phases in the periodicity conditions for the field operator $\psi ^{\prime
}(x)$ (in the following we will omit the primes) will be denoted by $\tilde{%
\alpha}_{l}$:
\begin{equation}
\tilde{\alpha}_{l}=\alpha _{l}+eA_{l}L_{l}.  \label{alft}
\end{equation}%
This shows that the physics depends on the parameters $\alpha _{l}$ and $%
A_{l}$ in the form of the combination (\ref{alft}). The phase shift induced
by the vector potential can be presented as $eA_{l}L_{l}=-eA^{l}L_{l}=-2\pi
\Phi _{l}/\Phi _{0}$, where $\Phi _{l}$ is formally interpreted in terms of
the magnetic flux enclosed by the $l$th compact dimension and $\Phi
_{0}=2\pi /e$ is the flux quantum.

In the presence of the branes at $z=z_{j}$, $j=1,2$, for the theory to be
defined one needs to specify the boundary conditions on them. In this
section we will assume that the field operator obeys the bag boundary
conditions
\begin{equation}
(1+i\gamma ^{\mu }n_{\mu }^{(j)})\psi (x)=0,\;z=z_{j},  \label{Bagbc}
\end{equation}%
with $n_{\mu }^{(j)}$ being the inward pointing normal (with respect to the
region under consideration) to the brane at $z=z_{j}$. Other types of
boundary conditions on the branes will be discussed in the following
sections. The branes divide the background space into three regions: $%
0\leqslant z\leqslant z_{1}$, $z_{1}\leqslant z\leqslant z_{2}$, and $%
z\geqslant z_{2}$. The current densities in the regions $0\leqslant
z\leqslant z_{1}$ and $z\geqslant z_{2}$ are the same as those for a single
brane located at $z=z_{1}$ and $z=z_{2}$, respectively, and they are
investigated in \cite{Bell18}. Here we will be mainly concerned with the
region between the branes, $z_{1}\leqslant z\leqslant z_{2}$. For that
region in (\ref{Bagbc}) one has $n_{\mu }^{(j)}=(-1)^{j}\delta _{\mu
}^{D}a/z_{j}$.

\section{Fermionic modes in the region between the branes}

\label{sec:Modes}

In this section we consider a complete set of positive and negative energy
modes $\{\psi _{\beta }^{(+)},\psi _{\beta }^{(-)}\}$ for the fermionic
field $\psi (x)$. The collective set $\beta $ of quantum numbers will be
specified below. In order to solve the field equation (\ref{Feq}) we need to
choose the representation of the Dirac matrices. As it already has been
discussed in \cite{Bell18}, it is convenient to take the flat spacetime
gamma matrices in the representation%
\begin{equation}
\gamma ^{(b)}=\left(
\begin{array}{cc}
0 & \chi _{b} \\
\left( -1\right) ^{1-\delta _{b}^{0}}\chi _{b}^{\dagger } & 0%
\end{array}%
\right) ,\;b=0,1,\ldots ,D-1,  \label{gamma}
\end{equation}%
and $\gamma ^{(D)}=si\,\mathrm{diag}(1,-1)$ with $s=\pm 1$. In
odd-dimensional spacetimes there exist two inequivalent irreducible
representations of the Clifford algebra and the values $s=+1$ and $s=-1$
correspond to those representations. In even spacetime dimensions the
irreducible representation of the Clifford algebra is unique, up to a
similarity transformation, and we can put $s=1$. For $D=2$ we can take $\chi
_{0}=\chi _{1}=1$ and the matrices $\gamma ^{(b)}$ are expressed in terms of
the Pauli matrices $\sigma _{\mathrm{P}\mu }$ as $\gamma ^{(0)}=\sigma _{%
\mathrm{P}1}$, $\gamma ^{(1)}=i\sigma _{\mathrm{P}2}$, $\gamma
^{(2)}=si\sigma _{\mathrm{P}3}$. The commutation relations for the $%
N/2\times N/2$ matrices $\chi _{b}$ are obtained from those for the Dirac
matrices $\gamma ^{(b)}$. They are reduced to $\chi _{0}^{\dagger }\chi
_{0}=1$, $\chi _{0}\chi _{b}^{\dagger }=\chi _{b}\chi _{0}^{\dagger }$, $%
\chi _{0}^{\dagger }\chi _{b}=\chi _{b}^{\dagger }\chi _{0}$, and%
\begin{eqnarray}
\chi _{b}\chi _{c}^{\dagger }+\chi _{c}\chi _{b}^{\dagger } &=&2\delta _{bc},
\notag \\
\chi _{b}^{\dagger }\chi _{c}+\chi _{c}^{\dagger }\chi _{b} &=&2\delta _{bc},
\label{xicom}
\end{eqnarray}%
where $b,c=1,2,\ldots ,D-1$.

With the curved spacetime gamma matrices $\gamma ^{\mu }=(z/a)\delta
_{b}^{\mu }\gamma ^{(b)}$, the complete set of solutions of the field
equation (\ref{Feq}) (with $A_{\mu }=0$ in the gauge under consideration)
can be found in a way similar to that given in \cite{Bell18}. Introducing
the one-column matrices $w^{(\sigma )}$, $\sigma =$ $1,\ldots ,N/2$, with $%
N/2$ rows and with the elements $w_{l}^{(\sigma )}=\delta _{l\sigma }$, for
the positive and negative energy mode functions one gets%
\begin{eqnarray}
\psi _{\beta }^{(+)}(x) &=&z^{\frac{D+1}{2}}e^{i\mathbf{kx}-i\omega t}\left(
\begin{array}{c}
\frac{\mathbf{k\chi }\chi _{0}^{\dagger }+i\lambda -\omega }{\omega }%
Z_{ma+s/2}(\lambda z)w^{(\sigma )} \\
i\chi _{0}^{\dagger }\frac{\mathbf{k\chi }\chi _{0}^{\dagger }+i\lambda
+\omega }{\omega }Z_{ma-s/2}(\lambda z)w^{(\sigma )}%
\end{array}%
\right) ,  \notag \\
\psi _{\beta }^{(-)}(x) &=&z^{\frac{D+1}{2}}e^{i\mathbf{kx}+i\omega t}\left(
\begin{array}{c}
i\chi _{0}\frac{\mathbf{k\chi }^{\dagger }\chi _{0}-i\lambda +\omega }{%
\omega }Z_{ma+s/2}(\lambda z)w^{(\sigma )} \\
\frac{\mathbf{k\chi }^{\dagger }\chi _{0}-i\lambda -\omega }{\omega }%
Z_{ma-s/2}(\lambda z)w^{(\sigma )}%
\end{array}%
\right) ,  \label{psiPM}
\end{eqnarray}%
where $\mathbf{k}=(k_{1},\ldots ,k_{D-1})$, $\omega =\sqrt{\lambda ^{2}+k^{2}%
}$, $k=|\mathbf{k}|$, $\mathbf{kx}=\sum_{l=1}^{D-1}k_{i}x^{i}$, and $\mathbf{%
k\chi }=\sum_{l=1}^{D-1}k_{l}\chi _{l}$. In (\ref{psiPM}), $Z_{\mu }(\lambda
z)$, with $\mu =ma\pm s/2$, is a cylinder function. We present it in the
form of a linear combination of the Bessel and Neumann functions $J_{\mu
}(\lambda z)$ and $Y_{\mu }(\lambda z)$:%
\begin{equation}
Z_{\mu }(\lambda z)=c_{1}J_{\mu }(\lambda z)+c_{2}Y_{\mu }(\lambda z),
\label{Z}
\end{equation}%
where the coefficients are determined by the normalization and boundary
conditions.

The momentum $\mathbf{k}$ in the mode functions (\ref{psiPM}) can be
decomposed as $\mathbf{k}=(\mathbf{k}_{(p)},\mathbf{k}_{(q)})$, where $%
\mathbf{k}_{(p)}=(k_{1},\ldots ,k_{p})$ and $\mathbf{k}_{(q)}=(k_{p+1},%
\ldots ,k_{D-1})$ correspond to the uncompact and compact subspaces. For the
components in the uncompact subspace, as usual, one has $-\infty
<k_{i}<+\infty $, $i=1,\ldots ,p$, whereas the components in the compact
subspace are discretized by the periodicity conditions (\ref{PCo}) (with the
replacement $\alpha _{l}\rightarrow \tilde{\alpha}_{l}$ in the gauge under
consideration) with the eigenvalues%
\begin{equation}
k_{l}=\frac{2\pi n_{l}+\tilde{\alpha}_{l}}{L_{l}},\;n_{l}=0,\pm 1,\pm
2,\ldots ,  \label{kl}
\end{equation}%
and $l=p+1,\ldots ,D-1$. In (\ref{kl}), the integer part of the ratio $%
\tilde{\alpha}_{l}/2\pi $ can be absorbed by the redefinition of $n_{l}$ and
only the fractional part of that ratio is physically relevant.

Now let us consider the boundary conditions on the branes. From (\ref{Bagbc}%
) it follows that $Z_{ma+1/2}(\lambda z_{1})=0$ and $Z_{ma-1/2}(\lambda
z_{2})=0$ in both the cases $s=\pm 1$. For the ratio of the coefficients in (%
\ref{Z}) the first condition gives
\begin{equation}
\frac{c_{2}}{c_{1}}=-\frac{J_{ma+1/2}(\lambda z_{1})}{Y_{ma+1/2}(\lambda
z_{1})}.  \label{crat}
\end{equation}%
From the second condition we obtain that the allowed values of $\lambda $
are roots of the equation%
\begin{equation}
g_{ma+1/2,ma-1/2}(\lambda z_{1},\lambda z_{2})=0,  \label{lamb}
\end{equation}%
where we have introduced the function
\begin{equation}
g_{\mu ,\nu }(x,u)=J_{\mu }(x)Y_{\nu }(u)-J_{\nu }(u)Y_{\mu }(x).
\label{gemu}
\end{equation}%
It can be seen that the equation (\ref{lamb}) has no solutions corresponding
to bound states with $\lambda =i\xi $, $\xi >0$. We denote the positive
roots of (\ref{lamb}) with respect to $\lambda z_{1}$ by $\lambda
_{n}=\lambda _{n}(ma,z_{2}/z_{1})=\lambda z_{1}$, $n=1,2,\ldots $, assuming
that they are numerated in the ascending order, $\lambda _{n+1}>\lambda _{n}$%
. Note that the roots $\lambda _{n}$ depend on the locations of the branes
in the form of the ratio $z_{2}/z_{1}$. For a massless field the equation (%
\ref{lamb}) is reduced to $\cos [\lambda \left( z_{2}-z_{1}\right) ]=0$ with
the solutions $\lambda z_{1}=\lambda _{n}=\pi (n-1/2)/\left(
z_{2}/z_{1}-1\right) $, $n=1,\ldots ,\infty $. These eigenvalues coincide
with those for parallel plates in the locally Minkowski bulk located at $%
z=z_{1}$ and $z=z_{2}$. For large values of $\lambda z_{1}$ and for a
massive field, in (\ref{lamb}) we can use the asymptotic expressions for the
cylinder functions for large arguments. To the leading order the left-hand
side of (\ref{lamb}) is reduced to $\cos [\lambda \left( z_{2}-z_{1}\right)
] $. Hence, for large $n$ one has $\lambda _{n}\approx \pi (n-1/2)/\left(
z_{2}/z_{1}-1\right) $. The mode functions are specified by the set $\beta =(%
\mathbf{k}_{(p)},\mathbf{n}_{q},n,\sigma )$ with $\mathbf{n}%
_{q}=(n_{p+1},\ldots ,n_{D-1})$.

The function $Z_{\nu }(\lambda z)$ is expressed in terms of the function (%
\ref{gemu}) in two equivalent ways:%
\begin{equation}
Z_{\mu }(\lambda z)=-\frac{c_{1}g_{ma+1/2,\mu }(\lambda z_{1},\lambda z)}{%
Y_{ma+1/2}(\lambda z_{1})}=-\frac{c_{1}g_{ma-1/2,\mu }(\lambda z_{2},\lambda
z)}{Y_{ma-1/2}(\lambda z_{2})}.  \label{Zg}
\end{equation}%
By using the first of these relations, and introducing the notation%
\begin{equation}
\nu =ma+1/2  \label{nu}
\end{equation}%
for the further convenience, the mode functions are presented in the form
\begin{eqnarray}
\psi _{\beta }^{(+)}(x) &=&C_{\beta }^{(+)}z^{\frac{D+1}{2}}e^{i\mathbf{kx}%
-i\omega t}\left(
\begin{array}{c}
\frac{\mathbf{k\chi }\chi _{0}^{\dagger }+i\lambda -\omega }{\omega }g_{\nu
,ma+s/2}(\lambda z_{1},\lambda z)w^{(\sigma )} \\
i\chi _{0}^{\dagger }\frac{\mathbf{k\chi }\chi _{0}^{\dagger }+i\lambda
+\omega }{\omega }g_{\nu ,ma-s/2}(\lambda z_{1},\lambda z)w^{(\sigma )}%
\end{array}%
\right) ,  \notag \\
\psi _{\beta }^{(-)}(x) &=&C_{\beta }^{(-)}z^{\frac{D+1}{2}}e^{i\mathbf{kx}%
+i\omega t}\left(
\begin{array}{c}
i\chi _{0}\frac{\mathbf{k\chi }^{\dagger }\chi _{0}-i\lambda +\omega }{%
\omega }g_{\nu ,ma+s/2}(\lambda z_{1},\lambda z)w^{(\sigma )} \\
\frac{\mathbf{k\chi }^{\dagger }\chi _{0}-i\lambda -\omega }{\omega }g_{\nu
,ma-s/2}(\lambda z_{1},\lambda z)w^{(\sigma )}%
\end{array}%
\right) ,  \label{psiPM2}
\end{eqnarray}%
with $\lambda =\lambda _{n}/z_{1}$ and new normalization coefficients $%
C_{\beta }^{(\pm )}$. The latter are found from the normalization condition
\begin{equation}
\int d^{D}x\,(a/z)^{D}\psi _{\beta }^{(\pm )\dagger }\psi _{\beta ^{\prime
}}^{(\pm )}=\delta _{\beta \beta ^{\prime }},  \label{nc}
\end{equation}%
where the $z$-integration goes over the range $[z_{1},z_{2}]$, $\delta
_{\beta \beta ^{\prime }}$ is understood as Kronecker delta for the discrete
components of the collective index $\beta $ and as Dirac delta function for
the continuous ones. The normalization integral over $z$ contains the
squares of the functions $g_{\nu ,ma\pm 1/2}(\lambda z_{1},\lambda z)$. By
taking into account that these functions are cylinder functions with respect
to both the arguments, the integrals are evaluated by using the
corresponding formula from \cite{Prud86}. Using the fact that $\lambda $ is
a root of the equation (\ref{lamb}), it can be shown that%
\begin{equation}
\int_{z_{1}}^{z_{2}}dz\,zg_{\nu ,ma\pm 1/2}^{2}(\lambda z_{1},\lambda z)=%
\frac{2\pi ^{-2}z_{1}/\lambda }{T_{\nu }(z_{2}/z_{1},\lambda z_{1})},
\label{Intg2}
\end{equation}%
where we have defined
\begin{equation}
T_{\nu }(\eta ,x)=x\left[ \frac{J_{\nu }^{2}(x)}{J_{\nu -1}^{2}(\eta x)}-1%
\right] ^{-1},\;\eta =z_{2}/z_{1}.  \label{Tnu}
\end{equation}%
On the base (\ref{Intg2}) it can be seen that
\begin{equation}
\left\vert C_{\beta }^{(\pm )}\right\vert ^{2}\equiv \left\vert C_{\beta
}\right\vert ^{2}=\frac{\lambda T_{\nu }(\eta ,\lambda z_{1})}{32(2\pi
)^{p-2}V_{q}a^{D}z_{1}}.  \label{Cbet}
\end{equation}%
As seen the normalization constants are the same (up to a phase) for the
positive and negative energy solutions and they do not depend on the
parameter $s$.

\section{Current density in the region between the branes}

\label{sec:Curr}

Having the complete set of normalized mode functions (\ref{psiPM2}), we can
evaluate the VEV of the current density $j^{\mu }=e\bar{\psi}\gamma ^{\mu
}\psi $, where $\bar{\psi}=\psi ^{\dagger }\gamma ^{(0)}$ is the Dirac
conjugate (for a recent discussion of the renormalised fermion expectation
values on AdS spacetime in the absence of branes see, for example, \cite%
{Ambr15}). That is done by using the mode-sum formula
\begin{equation}
\langle j^{\mu }(x)\rangle =\frac{e}{2}\sum_{\beta }\left[ \bar{\psi}_{\beta
}^{(-)}(x)\gamma ^{\mu }\psi _{\beta }^{(-)}(x)-\bar{\psi}_{\beta
}^{(+)}(x)\gamma ^{\mu }\psi _{\beta }^{(+)}(x)\right] ,  \label{jVEV}
\end{equation}%
where $\langle j^{\mu }(x)\rangle =\left\langle 0\right\vert j^{\mu
}(x)\left\vert 0\right\rangle $ with $\left\vert 0\right\rangle $ being the
vacuum state, and%
\begin{equation}
\sum_{\beta }=\sum_{\mathbf{n}_{q}}\int d\mathbf{k}_{(p)}\sum_{n=1}^{\infty
}\sum_{\sigma =1}^{N/2}\,.  \label{Sumbet}
\end{equation}%
We consider the charge density and the spatial components separately.

The component with $\mu =0$ in (\ref{jVEV}) corresponds to the VEV of the
charge density. Inserting the mode functions it is presented in the form
\begin{equation}
\langle j^{0}\rangle =\frac{(2\pi )^{2-p}ez^{D+2}}{16a^{D+1}V_{q}z_{1}}%
\sum_{\beta }\frac{\lambda }{\omega }T_{\nu }(\eta ,\lambda
z_{1})\sum_{j=\pm 1}jg_{\nu ,ma+js/2}^{2}(\lambda z_{1},\lambda z)w^{(\sigma
)\dagger }\mathbf{k\chi }^{\dagger }\chi _{0}w^{(\sigma )}.  \label{j0}
\end{equation}%
From the definition of the one-column matrices $w^{(\sigma )}$ it follows
that $\sum_{\sigma =1}^{N/2}w^{(\sigma )\dagger }\mathbf{\chi }^{\dagger
}\chi _{0}w^{(\sigma )}=\mathrm{tr}(\mathbf{\chi }^{\dagger }\chi _{0})$.
Now, by taking into account the commutation relations for the matrices $\chi
_{0}$ and $\chi _{b}$, we can show that $\mathrm{tr}(\chi _{b}^{\dagger
}\chi _{0})=0$ and consequently $\sum_{\sigma =1}^{N/2}w^{(\sigma )\dagger }%
\mathbf{\chi }^{\dagger }\chi _{0}w^{(\sigma )}=0$. Hence, we conclude that
the VEV\ of the charge density vanishes.

Next we consider the spatial components of the VEV (\ref{jVEV}). With the
mode functions (\ref{psiPM2}) we get
\begin{equation}
\langle j^{l}\rangle =-\frac{(2\pi )^{2-p}eNz^{D+2}}{32V_{q}a^{D+1}z_{1}}%
\sum_{\mathbf{n}_{q}}\int d\mathbf{k}_{(p)}k_{l}\sum_{n=1}^{\infty }\frac{%
\lambda }{\omega }T_{\nu }(\eta ,\lambda z_{1})\sum_{j=\pm 1}g_{\nu
,ma+js/2}^{2}(\lambda z_{1},\lambda z).  \label{jl}
\end{equation}%
For the components along uncompact dimensions, $l=1,\ldots ,p$, in (\ref{jl}%
) one has $-\infty <k_{l}<+\infty $ and in the integral $\int_{-\infty
}^{+\infty }dk_{l}$ the integrand is an odd function of the integration
variable. From here we conclude that the components of the current density
along uncompact dimensions vanish: $\langle j^{l}\rangle =0$ for $l=1,\ldots
,p$. This result could also be directly obtained on the base of the problem
symmetry under the reflections $x^{l}\rightarrow -x^{l}$ of the uncompact
directions. Hence, a nonzero vacuum currents may appear along the compact
dimensions only and we pass to the investigation of their properties.

\subsection{Integral representation for the currents in the compact subspace}

First of all, from (\ref{jl}) it follows that the current density does not
depend on the parameter $s$. This means that in odd-dimensional spacetimes
the current densities are the same for fermionic fields realizing two
inequivalent irreducible representations of the Clifford algebra. In the
following discussion we put $s=1$. After integrating over the angular part
of $\mathbf{k}_{(p)}$, the current density along the $l$th compact dimension
is presented in the form%
\begin{eqnarray}
\langle j^{l}\rangle &=&-\frac{\pi ^{2-p/2}eNz^{D+2}}{2^{p+2}\Gamma
(p/2)V_{q}a^{D+1}z_{1}}\sum_{\mathbf{n}_{q}}k_{l}\int_{0}^{\infty
}dk_{(p)}\,k_{(p)}^{p-1}  \notag \\
&&\times \sum_{n=1}^{\infty }\frac{\lambda _{n}T_{\nu }(\eta ,\lambda _{n})}{%
\sqrt{\lambda _{n}^{2}+z_{1}^{2}k^{2}}}\sum_{j=\pm 1}g_{\nu
,ma+j/2}^{2}(\lambda _{n},\lambda _{n}z/z_{1}),  \label{jl1}
\end{eqnarray}%
where $l=p+1,\ldots ,D-1$, $k^{2}=k_{(p)}^{2}+k_{(q)}^{2}$, $k_{(p)}^{2}=|%
\mathbf{k}_{(p)}|^{2}$ and%
\begin{equation}
k_{(q)}^{2}=\sum_{i=p+1}^{D-1}\frac{(2\pi n_{i}+\tilde{\alpha}_{i})^{2}}{%
L_{i}^{2}}.  \label{kq2}
\end{equation}%
From (\ref{jl1}) it follows that $\langle j^{l}\rangle $ is an odd periodic
function of $\tilde{\alpha}_{l}$ with the period $2\pi $ and an even
periodic function of $\tilde{\alpha}_{i}$, $i\neq l$, with the same period.
In terms of the magnetic fluxes $\Phi _{i}$ this means that the current
density is a periodic function of the magnetic fluxes with the period equal
to the flux quantum. In particular, $\langle j^{l}\rangle $ vanishes for
integer values of $\tilde{\alpha}_{l}/(2\pi )$. The charge flux density
through the hypersurface $x^{l}=\mathrm{const}$ is given by $%
n_{l}^{(l)}\langle j^{l}\rangle $, where $n_{i}^{(l)}=\delta _{i}^{l}a/z$ is
the normal to that hypersurface. The product $a^{D}n_{l}^{(l)}\langle
j^{l}\rangle $ depends on the variables having the dimension of length in
the form of the dimensionless combinations $z_{j}/z$, $L_{i}/z$, $ma$. This
feature is a consequence of the maximal symmetry of the AdS spacetime. In
figures below we plot the quantity $a^{D}n_{l}^{(l)}\langle j^{l}\rangle $.

In the representation (\ref{jl1}) the eigenvalues $\lambda _{n}$ are given
implicitly, as roots of the equation (\ref{lamb}). Another disadvantage is
that the terms with large $n$ are highly oscillatory. A more convenient
representation is obtained applying a variant of the generalized Abel-Plana
formula (\ref{SumFormAP}), derived in Appendix \ref{sec:App}, with $\mu =\nu
-1$ and $\delta =1$. Note that $\lambda _{\nu -1,n}^{(1)}=\lambda _{n}$ and $%
T_{\nu -1}^{(1)}(\eta ,x)=T_{\nu }(\eta ,x)$. For the series over $n$ in (%
\ref{jl1}) the function $h(u)$ has the form
\begin{equation}
h(x)=\frac{x}{\sqrt{x^{2}+z_{1}^{2}k^{2}}}\sum_{j=\pm 1}g_{\nu
,ma+j/2}^{2}(x,xz/z_{1}),  \label{hu}
\end{equation}%
and has branch points $x=\pm iz_{1}k$ on the imaginary axis. By using the
properties of the Bessel functions it can be seen that $h(ix)+h(-ix)=0$ for $%
0\leqslant x<z_{1}k$. With the help of (\ref{SumFormAP}) the current density
is presented as
\begin{eqnarray}
\langle j^{l}\rangle &=&\langle j^{l}\rangle ^{(1)}-\frac{4\left( 4\pi
\right) ^{-p/2-1}eNz^{D+2}}{\Gamma (p/2)V_{q}a^{D+1}z_{1}}\sum_{\mathbf{n}%
_{q}}k_{l}\int_{0}^{\infty }dk_{(p)}\,k_{(p)}^{p-1}  \notag \\
&&\times \int_{z_{1}k}^{\infty }dx\,x\frac{K_{\nu -1}(\eta x)}{K_{\nu }(x)}%
\frac{\sum_{j=\pm 1}jG_{\nu ,ma+j/2}^{2}(x,xz/z_{1})}{\sqrt{%
x^{2}-z_{1}^{2}k^{2}}G_{\nu ,\nu -1}(x,\eta x)},  \label{jl2}
\end{eqnarray}%
where we have defined
\begin{equation}
G_{\mu ,\nu }(x,u)=I_{\mu }(x)K_{\nu }(u)-(-1)^{\mu -\nu }K_{\mu }(x)I_{\nu
}(u).  \label{Gnumu}
\end{equation}%
with the modified Bessel functions $I_{\mu }(x)$ and $K_{\mu }(x)$. The
first term in the right-hand side of (\ref{jl2}) is given by%
\begin{eqnarray}
\langle j^{l}\rangle ^{(1)} &=&-\frac{\left( 4\pi \right) ^{-p/2}eNz^{D+2}}{%
2\Gamma (p/2)V_{q}a^{D+1}z_{1}}\sum_{\mathbf{n}_{q}}k_{l}\int_{0}^{\infty
}dk_{(p)}\,k_{(p)}^{p-1}  \notag \\
&&\times \int_{0}^{\infty }dx\frac{x}{\sqrt{x^{2}+z_{1}^{2}k^{2}}}\frac{%
\sum_{j=\pm 1}g_{\nu ,ma+j/2}^{2}(x,xz/z_{1})}{J_{\nu }^{2}(x)+Y_{\nu
}^{2}(x)},  \label{jl11}
\end{eqnarray}%
and it corresponds to the current density in the region $z_{1}\leqslant
z<\infty $ for the geometry of a single brane located at $z=z_{1}$.

The single brane part (\ref{jl11}) has been investigated in \cite{Bell18}.
It is decomposed as%
\begin{equation}
\langle j^{l}\rangle ^{(1)}=\langle j^{l}\rangle _{0}+\langle j^{l}\rangle
_{b}^{(1)},  \label{jdec}
\end{equation}%
where the term $\langle j^{l}\rangle _{0}$ is the current density in the
absence of the branes and
\begin{equation}
\langle j^{l}\rangle _{b}^{(1)}=\frac{NeA_{p}z^{D+2}}{V_{q}a^{D+1}}\sum_{%
\mathbf{n}_{q}}k_{l}\int_{k_{(q)}}^{\infty }du\,u(u^{2}-k_{(q)}^{2})^{\frac{%
p-1}{2}}\frac{I_{\nu }(z_{1}u)}{K_{\nu }(z_{1}u)}\sum_{j=\pm
1}jK_{ma+j/2}^{2}(zu),  \label{jlRb}
\end{equation}%
with
\begin{equation}
A_{p}=\frac{\left( 4\pi \right) ^{-(p+1)/2}}{\Gamma \left( (p+1)/2\right) },
\label{Cp}
\end{equation}%
is the part induced in the region $z\geqslant z_{1}$ by a single brane at $%
z=z_{1}$. The part $\langle j^{l}\rangle _{0}$ is investigated in \cite%
{Bell17}. It is expressed in terms of the function%
\begin{equation}
q_{\mu }^{\left( D+1\right) /2}(u)=\sqrt{\frac{\pi }{2}}\int_{0}^{\infty
}dx\,x^{D/2}e^{-ux}I_{\mu +1/2}(x),  \label{qf}
\end{equation}%
and is presented in the form
\begin{equation}
\langle j^{l}\rangle _{0}=-\frac{eNa^{-D-1}L_{l}}{(2\pi )^{(D+1)/2}}%
\sum_{n_{l}=1}^{\infty }n_{l}\sin (\tilde{\alpha}_{l}n_{l})\sum_{\mathbf{n}%
_{q-1}}\,\cos (\sum_{i=1,\neq l}^{D-1}\tilde{\alpha}_{i}n_{i})%
\sum_{j=0,1}q_{ma-j}^{\left( D+1\right) /2}\left( 1+\sum_{i=p+1}^{D-1}\frac{%
n_{i}^{2}L_{i}^{2}}{2z^{2}}\right) ,  \label{jl02}
\end{equation}%
where $\mathbf{n}_{q-1}=(n_{p+1},\ldots ,n_{l-1},n_{l+1},\ldots n_{D-1})$.
An alternative expression for the function (\ref{qf}) in terms of the
hypergeometric function is given in \cite{Bell17}.

The last term in (\ref{jl2}) is induced in the region $z_{1}\leqslant
z\leqslant z_{2}$ if we add to the geometry of a single brane at $z=z_{1}$
the second brane at $z=z_{2}$. It can be further transformed by introducing
a new integration variable $w=\sqrt{x^{2}-z_{1}^{2}k^{2}}$ and passing to
polar coordinates $(r,\theta )$ in the plane $(z_{1}k_{(p)},w)$. After
integrating over $\theta $ and introducing instead of $r$ the integration
variable $u=\sqrt{r^{2}/z_{1}^{2}+k_{(q)}^{2}}$, we find%
\begin{eqnarray}
\langle j^{l}\rangle &=&\langle j^{l}\rangle _{0}+\langle j^{l}\rangle
_{b}^{(1)}-\frac{NeA_{p}z^{D+2}}{V_{q}a^{D+1}}\sum_{\mathbf{n}%
_{q}}k_{l}\int_{k_{(q)}}^{\infty }du\,u\frac{K_{\nu -1}(z_{2}u)}{K_{\nu
}(z_{1}u)}  \notag \\
&&\times \frac{(u^{2}-k_{(q)}^{2})^{\frac{p-1}{2}}}{G_{\nu ,\nu
-1}(z_{1}u,z_{2}u)}\sum_{j=\pm 1}jG_{\nu ,ma+j/2}^{2}(z_{1}u,zu).
\label{jl4}
\end{eqnarray}%
By taking into account the representation (\ref{jdec}) for a single brane
part, the current density is presented as%
\begin{eqnarray}
\langle j^{l}\rangle &=&\langle j^{l}\rangle _{0}+\frac{NeA_{p}z^{D+2}}{%
V_{q}a^{D+1}}\sum_{\mathbf{n}_{q}}k_{l}\int_{k_{(q)}}^{\infty
}du\,u(u^{2}-k_{(q)}^{2})^{\frac{p-1}{2}}\left[ \frac{K_{\nu }(z_{1}u)I_{\nu
-1}(z_{2}u)}{I_{\nu }(z_{1}u)K_{\nu -1}(z_{2}u)}+1\right] ^{-1}  \notag \\
&&\times \sum_{j=\pm 1}j\left[ \frac{I_{\nu -1}(z_{2}u)}{K_{\nu -1}(z_{2}u)}%
K_{ma+j/2}^{2}(zu)-\frac{K_{\nu }(z_{1}u)}{I_{\nu }(z_{1}u)}%
I_{ma+j/2}^{2}(zu)\right.  \notag \\
&&\left. +2jK_{ma+j/2}(zu)I_{ma+j/2}(zu)\right] .  \label{jl5}
\end{eqnarray}%
An alternative representation is given by the formula%
\begin{eqnarray}
\langle j^{l}\rangle &=&\langle j^{l}\rangle _{0}+\langle j^{l}\rangle
_{b}^{(2)}+\frac{NeA_{p}z^{D+2}}{V_{q}a^{D+1}}\sum_{\mathbf{n}%
_{q}}k_{l}\int_{k_{(q)}}^{\infty }du\,u\frac{I_{\nu }(z_{1}u)}{I_{\nu
-1}(z_{2}u)}  \notag \\
&&\times \frac{(u^{2}-k_{(q)}^{2})^{\frac{p-1}{2}}}{G_{\nu ,\nu
-1}(z_{1}u,z_{2}u)}\sum_{j=\pm 1}jG_{\nu -1,ma+j/2}^{2}(z_{2}u,zu),
\label{jl6}
\end{eqnarray}%
where
\begin{equation}
\langle j^{l}\rangle _{b}^{(2)}=-\frac{NeA_{p}z^{D+2}}{V_{q}a^{D+1}}\sum_{%
\mathbf{n}_{q}}k_{l}\int_{k_{(q)}}^{\infty }du\,u(u^{2}-k_{(q)}^{2})^{\frac{%
p-1}{2}}\frac{K_{\nu -1}(uz_{2})}{I_{\nu -1}(uz_{2})}\sum_{j=\pm
1}jI_{ma+j/2}^{2}(uz),  \label{jb2}
\end{equation}%
is the current density induced by a single brane at $z=z_{2}$ in the region $%
0\leqslant z\leqslant z_{2}$ (see \cite{Bell18}). The last term in (\ref{jl6}%
) is induced by the brane at $z=z_{1}$ if we add it to the problem with a
single brane at $z=z_{2}$. As it will be seen in the next subsection, the
VEV of the current density is finite on the branes. However, in the
representations (\ref{jlRb}) and (\ref{jb2}) for single brane-induced parts
we cannot directly put in the integrands $z=z_{1}$ and $z=z_{2}$,
respectively. This can be done in the second brane-induced contributions
(last terms in (\ref{jl4}) and (\ref{jl6})).

In order to see the interference effects between the branes we can present
the total current density as%
\begin{equation}
\langle j^{l}\rangle =\langle j^{l}\rangle _{0}+\langle j^{l}\rangle
_{b}^{(1)}+\langle j^{l}\rangle _{b}^{(2)}+\langle j^{l}\rangle _{\mathrm{int%
}}.  \label{jdecint}
\end{equation}%
By taking into account the expressions for the single brane parts, for the
interference part we can get the expression%
\begin{eqnarray}
\langle j^{l}\rangle _{\mathrm{int}} &=&\frac{NeA_{p}z^{D+2}}{V_{q}a^{D+1}}%
\sum_{\mathbf{n}_{q}}k_{l}\int_{k_{(q)}}^{\infty }du\,u(u^{2}-k_{(q)}^{2})^{%
\frac{p-1}{2}}\left[ \frac{K_{\nu }(z_{1}u)I_{\nu -1}(z_{2}u)}{I_{\nu
}(z_{1}u)K_{\nu -1}(z_{2}u)}+1\right] ^{-1}  \notag \\
&&\times \sum_{j=\pm 1}j\left[ \frac{K_{\nu -1}(z_{2}u)}{I_{\nu -1}(z_{2}u)}%
I_{ma+j/2}^{2}(zu)-\frac{I_{\nu }(z_{1}u)}{K_{\nu }(z_{1}u)}%
K_{ma+j/2}^{2}(zu)\right.  \notag \\
&&\left. +2jK_{ma+j/2}(zu)I_{ma+j/2}(zu)\right] .  \label{jint}
\end{eqnarray}%
Note that in the evaluation of the interference part on the branes we can
directly put in the integrand $z=z_{j}$.

The current densities in the regions $z\leqslant z_{1}$ and $z\geqslant
z_{2} $ coincide with those in the corresponding geometries with single
branes. In these regions the VEV\ is presented as $\langle j^{l}\rangle
=\langle j^{l}\rangle _{0}+\langle j^{l}\rangle _{b}$, where the
brane-induced contribution $\langle j^{l}\rangle _{b}$ is given by (\ref{jb2}%
) in the region $z\leqslant z_{1}$, with the replacement $z_{2}\rightarrow
z_{1}$, and by (\ref{jlRb}) in the region $z\geqslant z_{2}$, with the
replacement $z_{1}\rightarrow z_{2}$.

\subsection{Alternative representation and the currents on the branes}

Here we provide another representation for the VEV\ of the current density
that is more adapted for the investigation of the near-brane asymptotics. It
is obtained from the initial expression (\ref{jl1}) by using the summation
formula \cite{Bell10}%
\begin{eqnarray}
&&\frac{2\pi }{L_{l}}\sum_{n_{l}=-\infty }^{\infty
}g(k_{l})f(|k_{l}|)=\int_{0}^{\infty }du[g(u)+g(-u)]f(u)  \notag \\
&&\qquad +i\int_{0}^{\infty }du\,[f(iu)-f(-iu)]\sum_{j=\pm 1}\frac{g(iju)}{%
e^{uL_{l}+ij\tilde{\alpha}_{l}}-1},  \label{apnl}
\end{eqnarray}%
where $k_{l}$ is given by (\ref{kl}). In the special case $g(x)=1$, $\tilde{%
\alpha}_{l}=0$ the standard Abel-Plana formula is obtained from (\ref{apnl}%
). For the series over $n_{l}$ in (\ref{jl1}) we have $g(u)=u$ and the first
integral in the right-hand side of (\ref{apnl}) vanishes. Physically this
corresponds to the fact that the part in the current density with that
integral presents the current in the model where the dimension $x^{l}$ is
decompactified and, hence, as it has been shown above, the corresponding
current is zero. With $g(u)=u$, by using the expansion $1/(e^{y}-1)=%
\sum_{r=1}^{\infty }e^{-ry}$, after evaluating the integrals over $u$ and $%
k_{(p)}$, one gets
\begin{eqnarray}
\langle j^{l}\rangle &=&-\frac{\left( 2\pi \right) ^{1-p/2}eNz^{D+2}}{%
8V_{q}L_{l}^{p}a^{D+1}z_{1}^{2}}\sum_{r=1}^{\infty }\frac{\sin (r\tilde{%
\alpha}_{l})}{r^{p+1}}\sum_{\mathbf{n}_{q-1}}\sum_{n=1}^{\infty }\lambda
_{n}T_{\nu }(\eta ,\lambda _{n})  \notag \\
&&\times g_{p/2+1}(rL_{l}\sqrt{\lambda _{n}^{2}/z_{1}^{2}+k_{(q-1)}^{2}}%
)\sum_{j=\pm 1}g_{\nu ,ma+j/2}^{2}(\lambda _{n},\lambda _{n}z/z_{1}).
\label{jl7}
\end{eqnarray}%
where $k_{(q-1)}^{2}=\sum_{i=p+1,\neq l}^{D-1}\left( 2\pi n_{i}+\tilde{\alpha%
}_{i}\right) ^{2}/L_{i}^{2}$ and we have defined the function
\begin{equation}
g_{\mu }(x)=x^{\mu }K_{\mu }(x).  \label{gen}
\end{equation}%
Unlike to the series over $n$ in (\ref{jl1}) the corresponding series in (%
\ref{jl7}) is exponentially convergent.

The representation (\ref{jl7}) is well adapted for the investigation of the
currents on the branes. They are obtained putting $z=z_{j}$ directly in the
right-hand side of (\ref{jl7}). By taking into account that $g_{\nu ,\nu
-1}(\lambda _{n},\eta \lambda _{n})=0$, and%
\begin{eqnarray}
g_{\nu ,\nu -1}(\lambda _{n},\lambda _{n}) &=&\frac{2}{\pi \lambda _{n}},
\notag \\
g_{\nu ,\nu }(\lambda _{n},\eta \lambda _{n}) &=&-\frac{2}{\pi \eta \lambda
_{n}}\frac{J_{\nu }(\lambda _{n})}{J_{\nu -1}(\eta \lambda _{n})},
\label{jbr}
\end{eqnarray}%
one gets%
\begin{eqnarray}
\langle j^{l}\rangle _{z=z_{j}} &=&-\frac{2eNL_{l}^{-p}z_{j}^{D}}{\left(
2\pi \right) ^{p/2+1}V_{q}a^{D+1}}\sum_{r=1}^{\infty }\frac{\sin (r\tilde{%
\alpha}_{l})}{r^{p+1}}\sum_{n=1}^{\infty }\left[ \frac{J_{\nu }(\lambda _{n})%
}{J_{\nu -1}(\eta \lambda _{n})}\right] ^{2(j-1)}  \notag \\
&&\times \sum_{\mathbf{n}_{q-1}}\frac{g_{p/2+1}(rL_{l}\sqrt{\lambda
_{n}^{2}/z_{1}^{2}+k_{(q-1)}^{2}})}{J_{\nu }^{2}(\lambda _{n})/J_{\nu
-1}^{2}(\eta \lambda _{n})-1}.  \label{jlzj}
\end{eqnarray}%
Finiteness of the vacuum current density on the branes is in clear contrast
with the corresponding behavior of the fermionic condensate and of the VEV
of the energy-momentum tensor. The latter VEVs diverge on the boundaries.
This kind of surface divergences have been widely discussed in the Casimir
effect for fields with different spins and for different boundary
geometries. The absence of the surface divergences for the current density
in the problem under consideration can be understood from general arguments.
In the problem with two branes and without compact dimensions the VEV of the
current density vanishes. The compactification scheme we cave considered
does not change the local bulk and boundary geometries. By taking into
account that the divergences are completely determined by those local
geometries, we conclude that the toral compactification will not induce
additional divergences in the VEVs. In particular, the VEV of the current
density becomes finite everywhere.

We can also use the representation (\ref{jl7}) for the evaluation of the
total current, per unit surface along the uncompact dimensions. By using the
integrals (\ref{Intg2}) we get%
\begin{eqnarray}
V_{q}\int_{z_{1}}^{z_{2}}dz\,\sqrt{|g|}\langle j^{l}\rangle &=&-\frac{2Ne}{%
(2\pi )^{p/2+1}L_{l}^{p}}\sum_{r=1}^{\infty }\frac{\sin (r\tilde{\alpha}_{l})%
}{r^{p+1}}  \notag \\
&&\times \sum_{\mathbf{n}_{q-1}}\sum_{n=1}^{\infty }g_{p/2+1}(rL_{l}\sqrt{%
\lambda _{n}^{2}/z_{1}^{2}+k_{(q-1)}^{2}}),  \label{jlint}
\end{eqnarray}%
where $g$ is the determinant of the metric tensor. In (\ref{jlint}), the
information on the curvature and on the boundary geometry is encoded through
the ratio $\lambda _{n}/z_{1}$. For a given distance between the branes the
ratio $z_{2}/z_{1}$ is fixed and the roots $\lambda _{n}$ do not depend on
the location of the left brane $z_{1}$. In particular, from here it follows
that, for fixed $z_{2}/z_{1}$, the quantity (\ref{jlint}) goes to zero in
the limit $z_{1}\rightarrow 0$. Comparing the integrated current (\ref{jlint}%
) with the current densities (\ref{jlzj}) on the branes, the following
simple relation between them is obtained.
\begin{equation}
\int_{z_{1}}^{z_{2}}dz\,\sqrt{|g|}\langle j^{l}\rangle =\left[ \frac{a^{D+1}%
}{z^{D}}\langle j^{l}\rangle \right] _{z=z_{1}}^{z=z_{2}}.  \label{relint}
\end{equation}

In the model with a single compact dimension $x^{l}$ with the length $L_{l}$
($q=1$, $p=D-2$, $l=D-1$) the formula (\ref{jl7}) is specified to%
\begin{eqnarray}
\langle j^{l}\rangle &=&-\frac{\left( 2\pi \right) ^{-D/2}eNz^{D+2}}{%
8L_{l}^{D-1}a^{D+1}z_{1}^{2}}\sum_{r=1}^{\infty }\frac{\sin (r\tilde{\alpha}%
_{l})}{r^{D-1}}\sum_{n=1}^{\infty }\lambda _{n}T_{\nu }(\eta ,\lambda _{n})
\notag \\
&&\times g_{D/2}(rL_{l}\lambda _{n}/z_{1})\sum_{j=\pm 1}g_{\nu
,ma+j/2}^{2}(\lambda _{n},\lambda _{n}z/z_{1}).  \label{jlq11}
\end{eqnarray}%
An alternative expression in this special case is obtained from (\ref{jl5}).
In this and in the next sections, for numerical investigations of the
current density we consider the special case $D=4$ with a single compact
dimension of the length $L_{l}=L$ and with the phase in the periodicity
condition $\tilde{\alpha}_{l}=\tilde{\alpha}$. For this model the
corresponding formulas are obtained from (\ref{jl5}) and (\ref{jlq11})
taking $p=2$ and $q=1$.

\subsection{Asymptotics and numerical examples}

In this subsection we consider the behavior of the current density in
asymptotic regions of the parameters. The Minkowskian limit corresponds to $%
a\rightarrow \infty $ for fixed $y$ and $y_{j}$. In this limit the conformal
coordinates $z$ and $z_{j}$ are large, $z\approx a+y$, $z-z_{j}\approx
y-y_{j}$, and, consequently, both the order and the argument of the modified
Bessel functions in (\ref{jl5}) are large. By using the corresponding
uniform asymptotic expansions \cite{Abra72}, for the brane-induced part, to
the leading order, we get $\langle j^{l}\rangle -\langle j^{l}\rangle
_{0}\approx $ $\langle j^{l}\rangle _{b}^{(M)}$, where%
\begin{eqnarray}
\langle j^{l}\rangle _{b}^{(M)} &=&\frac{2NeA_{p}}{V_{q}}\sum_{\mathbf{n}%
_{q}}k_{l}\int_{m_{(q)}}^{\infty }dx\,\frac{(x^{2}-m_{(q)}^{2})^{\frac{p-1}{2%
}}}{\frac{x+m}{x-m}e^{2x\left( y_{2}-y_{1}\right) }+1}  \notag \\
&&\times \left\{ 1+\frac{me^{x\left( y_{2}-y_{1}\right) }}{x-m}\cosh \left[
x\left( 2y-y_{2}-y_{1}\right) \right] \right\} ,  \label{jlbM}
\end{eqnarray}%
with the notation $m_{(q)}=\sqrt{m^{2}+k_{(q)}^{2}}$. This expression
coincides with the result from \cite{Bell13} for two boundaries in a flat
bulk with topology $R^{p+1}\times T^{q}$ (with the sign difference related
to definition of the parameters $\tilde{\alpha}_{i}$).

For a massless fermionic field the modified Bessel functions in (\ref{jl4})
are expressed in terms of the elementary functions. By taking into account
the expressions
\begin{eqnarray}
G_{1/2,1/2}(x,y) &=&\frac{\sinh (x-y)}{\sqrt{xy}},  \notag \\
G_{1/2,-1/2}(x,y) &=&\frac{\cosh (x-y)}{\sqrt{xy}},  \label{G12}
\end{eqnarray}%
we can see that%
\begin{equation}
\langle j^{l}\rangle =\langle j^{l}\rangle _{0}+\frac{NeA_{p}}{V_{q}}%
(z/a)^{D+1}\sum_{\mathbf{n}_{q}}k_{l}\int_{k_{(q)}}^{\infty }du\,\frac{%
(u^{2}-k_{(q)}^{2})^{\frac{p-1}{2}}}{e^{2u(z_{2}-z_{1})}+1}.  \label{jlm0}
\end{equation}%
The massless fermionic field is conformally invariant and, as we could
expect, the brane-induced contribution in (\ref{jlm0}) is conformally
related to the corresponding expression for two parallel boundaries in the
Minkowski bulk. The latter is obtained from (\ref{jlbM}) taking $m=0$.

Now let us consider the asymtotics for limiting cases of the brane
locations. In the limit $z_{2}\rightarrow \infty $, for fixed $z_{1}$ and $z$%
, the right brane tends to the AdS horizon. In this limit, it is expected
that from the results given above the current density will be obtained in
the region $z_{1}\leqslant z<\infty $ for the geometry of a single brane at $%
z=z_{1}$. In order to show that we use the representation (\ref{jl4}). The
part with $\langle j^{l}\rangle ^{(1)}$ does not depend on $z_{2}$ and it is
sufficient to consider the limiting transition for the last term. The latter
presents the contribution induced by the right brane. The dominant
contribution comes from the region of the integration near the lower limit
and from the mode in the summation over $\mathbf{n}_{q}$\ with the smallest
value of $k_{(q)}$. Under the assumption $|\tilde{\alpha}_{i}|<\pi $ that
mode corresponds to $n_{i}=0$ for $i=p+1,\ldots ,D-1$, and the corresponding
value for $k_{(q)}$ is given by
\begin{equation}
k_{(q)}^{(0)2}=\sum_{i=p+1}^{D-1}\tilde{\alpha}_{i}^{2}/L_{i}^{2}.
\label{kq0}
\end{equation}%
Hence, to the leading order, we get%
\begin{equation}
\langle j^{l}\rangle \approx \langle j^{l}\rangle ^{(1)}-\frac{\pi
Nek_{(q)}^{(0)(p+1)/2}\tilde{\alpha}_{l}z^{D+2}e^{-2z_{2}k_{(q)}^{(0)}}}{%
2V_{q}L_{l}a^{D+1}\left( 4\pi z_{2}\right) ^{(p+1)/2}}\sum_{j=\pm 1}j\frac{%
G_{\nu ,\nu _{j}}^{2}(z_{1}k_{(q)}^{(0)},zk_{(q)}^{(0)})}{K_{\nu
}^{2}(z_{1}k_{(q)}^{(0)})}.  \label{jlLargez2}
\end{equation}%
This shows that when the right brane tends to the AdS horizon the
corresponding contribution in the VEV of the current density is suppressed
by the factor $e^{-2z_{2}k_{(q)}^{(0)}}/z_{2}^{(p+1)/2}$.

In the limit $z_{1}\rightarrow 0$, for fixed $z_{2}$ and $z$, the left brane
tends to the AdS boundary. We use the representation (\ref{jl6}), where the
contribution of the left brane is given by the last term. To the leading
order, that contribution is obtained by using the asymptotic expressions of
the modified Bessel functions for small values of the arguments. In this way
it can be seen that in the limit when the left brane tends to the AdS
boundary the corresponding contribution to the current density vanishes as $%
z_{1}^{2ma+1}$.

Now let us consider the asymptotics with respect to the lengths of compact
dimensions. First let us discuss the case $L_{l}\ll L_{i}$, $i\neq l$. In
this limit the contribution of the modes with large $|n_{i}|$, $i\neq l$,
dominates in the VEV (\ref{jl5}) and, to the leading order, the
corresponding summations over $\mathbf{n}_{q-1}$ can be replaced by the
integration:%
\begin{equation}
\sum_{\mathbf{n}_{q-1}}f(k_{(q-1)})\rightarrow \frac{2\left( 4\pi \right)
^{(1-q)/2}V_{q}}{\Gamma ((q-1)/2)L_{l}}\int dx\,x^{q-2}\,f(x).  \label{Rep}
\end{equation}%
Next, we introduce a new integration variable $w=\sqrt{u^{2}-x^{2}-k_{l}^{2}}
$ and then pass to polar coordinates in the $(x,w)$-plane. After integrating
over the angular part one can see that, in the leading order, the current
density is obtained in a $(D+1)$-dimensional model with a single compact
dimension $x^{l}$: $\langle j^{l}\rangle \approx \langle j^{l}\rangle
|_{q=1} $. If additionally one has $L_{l}\ll z_{1}$, we can replace the
modified Bessel functions by the corresponding asymptotic expressions for
large arguments:%
\begin{equation}
\langle j^{l}\rangle \approx \langle j^{l}\rangle _{0}+\frac{2(4\pi
)^{(1-D)/2}Ne(z/a)^{D+1}}{\Gamma ((D-1)/2)L_{l}}\sum_{n_{l}=-\infty
}^{+\infty }k_{l}\int_{|k_{l}|}^{\infty }du\,\frac{\left(
u^{2}-k_{l}^{2}\right) ^{\frac{D-3}{2}}}{e^{2(z_{2}-z_{1})u}+1}.
\label{jlsmL}
\end{equation}%
Comparing with (\ref{jlbM}), we see that the brane-induced contribution in (%
\ref{jlsmL}) is conformally related to the corresponding current density for
a massless fermionic field in $(D+1)$-dimensional Minkowski spacetime with a
single compact dimension and with two planar boundaries having the distance $%
z_{2}-z_{1}$. In this limit the effects of the gravitational field are weak.
Under the additional constraint $L_{l}\ll (z_{2}-z_{1})$, the exponent in (%
\ref{jlsmL}) is large and we can further simplify the corresponding
expression. By taking into account that the dominant contribution comes from
$k_{l}$ with the minimal value $|k_{l}|$, we get%
\begin{equation}
\langle j^{l}\rangle \approx \langle j^{l}\rangle _{0}+\frac{\pi
^{(1-D)/2}Ne(z/a)^{D+1}\tilde{\alpha}_{l}|\tilde{\alpha}_{l}|^{\frac{D-3}{2}}%
}{2^{D-1}L_{l}^{(D+1)/2}(z_{2}-z_{1})^{(D-1)/2}}e^{-2(z_{2}-z_{1})|\tilde{%
\alpha}_{l}|/L_{l}},  \label{jlsmLb}
\end{equation}%
where it is assumed that $|\tilde{\alpha}_{l}|<\pi $. As seen, the
brane-induced contribution is exponentially small. Note that in the same
limit, $L_{l}\ll L_{i}$, $i\neq l$, and $L_{l}\ll z$, for the brane-free
contribution one has \cite{Bell17}%
\begin{equation}
\langle j^{l}\rangle _{0}\approx -\frac{eNL_{l}\Gamma ((D+1)/2)}{\pi
^{(D+1)/2}\left( aL_{l}/z\right) ^{D+1}}\sum_{n_{l}=1}^{\infty }\frac{\sin (%
\tilde{\alpha}_{l}n_{l})}{n_{l}^{D}}.  \label{jlLsm}
\end{equation}%
and it dominates in the total current density.

For large values of $L_{l}\gg L_{i},z_{1}$, $i\neq l$, it is more convenient
to use the representation (\ref{jl7}). The current density is dominated by
the lowest mode for $\lambda _{n}$ and by the mode for which $k_{(q-1)}^{2}$
takes its minimal value. For $|\tilde{\alpha}_{i}|<\pi $ the latter
corresponds to the mode with $n_{i}=0$, $i\neq l$, with the minimal value $%
k_{(q-1)}^{(0)2}=\sum_{i=p+1,\neq l}^{D-1}\tilde{\alpha}_{i}^{2}/L_{i}^{2}$.
By using the asymptotic expression of the Macdonald function for large
arguments, we can see that in the limit under consideration the current
density is suppressed by the factor $\exp [-L_{l}\sqrt{\lambda
_{1}^{2}/z_{1}^{2}+k_{(q-1)}^{(0)2}}]$.

If the one of the lengths $L_{i}$, $i\neq l$, is large compared to the other
length scales in the model, the expression (\ref{jl5}) for the current
density is dominated by the terms with large values of $|n_{i}|$. To the
leading order, we replace the corresponding summation by the integration and
the VEV $\langle j^{l}\rangle $ coincides with the current density in the
same model with decompactified $i$th coordinate $x^{i}$. In the opposite
limit of small $L_{i}$, , $i\neq l$, assuming that $|\tilde{\alpha}_{i}|<\pi
$, the behavior of the $l$th component of the current density crucially
depends on that wether $\tilde{\alpha}_{i}$ is zero or not. For $\tilde{%
\alpha}_{i}=0$ there is a zero mode along the $i$th compact dimension with $%
n_{i}=0$ and it dominates in the current density $\langle j^{l}\rangle $.
The leading term is obtained from (\ref{jl5}) taking the contribution with $%
n_{i}=0$ and we get $\langle j^{l}\rangle \approx Nz\langle j^{l}\rangle
_{D}/(N_{D}aL_{i})$, where $\langle j^{l}\rangle _{D}$ is the current
density in the $D$-dimensional model with the absence of the $i$th compact
dimension, $N_{D}$ is the number of spinor components in that model. For $%
\tilde{\alpha}_{i}\neq 0$, again, the dominant contribution comes from the
mode with $n_{i}=0$. The corresponding estimates can be done in a way
similar to that for small values of $L_{l}$ and we can see that the
brane-induced VEV is suppressed by the factor $e^{-2(z_{2}-z_{1})|\tilde{%
\alpha}_{i}|/L_{i}}$.

In the numerical examples of this section we will consider the model $D=4$
with a single compact dimension $x^{D}$. For the corresponding values of the
parameters one has $p=2$ and $q=1$. The length of the compact dimension will
be denoted by $L$ and the corresponding phase by $\tilde{\alpha}$. Four
different types of boundary conditions on the branes will be discussed
(corresponding to roman numerals near the graphs). Graphs with I correspond
to the bag boundary condition (\ref{Bagbc}) and the graphs with II
correspond to the condition (\ref{BC2}) below. As it has been discussed in
Section \ref{sec:Brane}, depending on the parity of the field under the
reflections with respect to the branes, two other classes of boundary
condition may arise in in $Z_{2}$-symmetric braneworld models. They
correspond to the boundary conditions $Z_{ma+1/2}(\lambda z_{j})=0$ (the
graphs will be designated by III) and $Z_{ma-1/2}(\lambda z_{j})=0$ (the
graphs designated by IV) on both the branes $z=z_{j}$, $j=1,2$.

\begin{figure}[tbph]
\begin{center}
\epsfig{figure=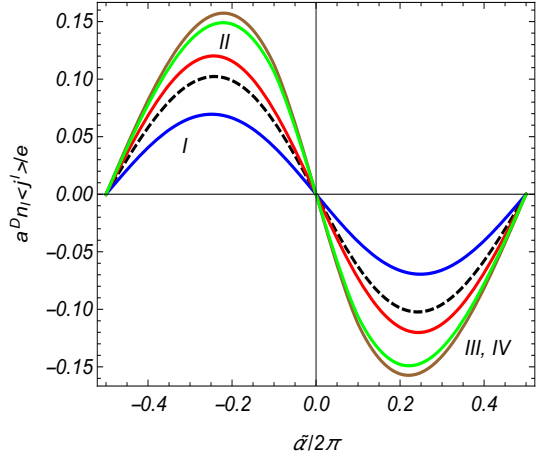,width=7.cm,height=6.cm}
\end{center}
\caption{Fermionic current along the compact dimension as a function of the
phase in the periodicity condition in the model $D=4$, $p=2$, $q=1$. The
roman numerals near the curves correspond to different classes of boundary
conditions on the branes. The graphs are plotted for $z_{1}/L=0.5$, $%
z_{2}/L=1$, $z/L=0.75$, and $ma=1$. }
\label{fig1}
\end{figure}

In figure \ref{fig1} we have displayed the dependence of the current density
on the phase $\tilde{\alpha}$. The current density is a periodic function of
$\tilde{\alpha}$ and graphs are plotted for one period. For the parameters
we have taken the values corresponding to $ma=1$, $z_{1}/L=0.5$, $z_{2}/L=1$%
, $z/L=0.75$. The dashed line corresponds to the current density in the
geometry without branes. As seen, depending on the boundary conditions
imposed on the field, the presence of the branes can either increase or
decrease the vacuum current density. In particular, the bag boundary
condition reduces the current density.

\begin{figure}[tbph]
\begin{center}
\epsfig{figure=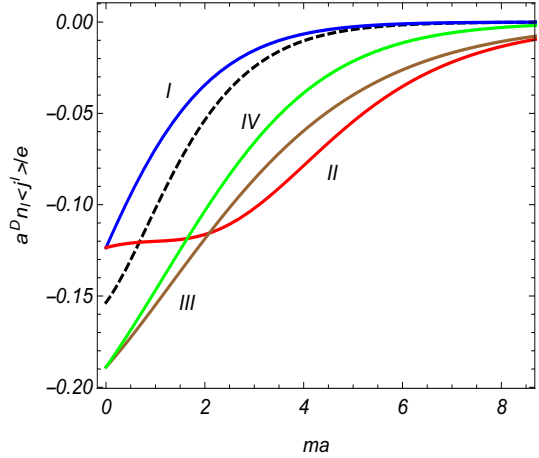,width=7.cm,height=6.cm}
\end{center}
\caption{The current density as a function of the field mass for $\tilde{%
\protect\alpha}=\protect\pi /2$. The other parameters are the same as those
for figure \protect\ref{fig1}.}
\label{fig2}
\end{figure}

The dependence of the current density on the field mass for different types
of boundary conditions is presented in figure \ref{fig2}. For the phase in
the periodicity condition we have taken $\tilde{\alpha}=\pi /2$. The values
of the remaining parameters are the same as those for figure \ref{fig1}.
From the data in figure \ref{fig2} we see that in the range of the mass $%
ma>1 $ the brane-induced currents can be essentially larger compared with
the currents in the brane-free geometry. Of course, for $ma\gg 1$ both these
contributions are suppressed. The coincidence of the current densities for a
massless field in the cases of boundary conditions I,II and III,IV will be
explained below on the base of the corresponding analytic expressions.

\begin{figure}[tbph]
\begin{center}
\epsfig{figure=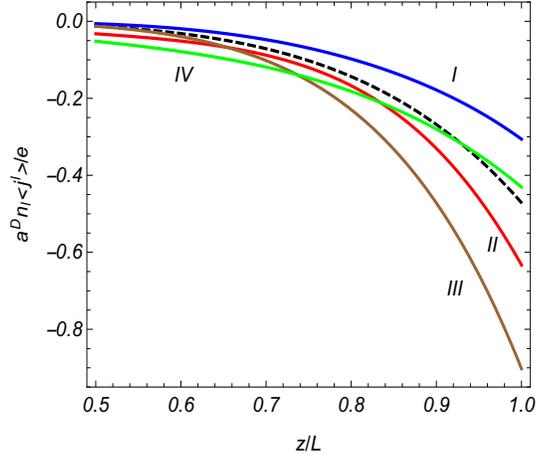,width=7.cm,height=6.cm}
\end{center}
\caption{The current density in the region between the branes as a function
of the coordinate $z$ in units of the length of the compact dimension. The
dashed curve presents the current density in the absence of the branes.}
\label{fig3}
\end{figure}

The behavior of the current density versus the coordinate $z$ is shown in
figure \ref{fig3} for $\tilde{\alpha}=\pi /2$, $ma=1$ and for the locations
of the branes we have taken $z_{1}/L=0.5$, $z_{2}/L=1$. As seen, the current
density is mainly located near the right brane.

It is also of interest to consider the dependence of the current density on
the length of the compact dimension. As it has been shown by the asymptotic
analysis that dependence is essentially different for the brane-free and
brane-induced contributions. For small values of $L$ the brane-free part
behaves as $\langle j^{l}\rangle _{0}\propto 1/\left( aL/z\right) ^{D+1}$
and the brane-induced part is suppressed by the factor $e^{-2(z_{2}-z_{1})|%
\tilde{\alpha}|/L}$ (see (\ref{jlsmLb})). This feature is seen in figure \ref%
{fig4}, where we have plotted the brane-induced current density, $\langle
j^{l}\rangle _{b}=\langle j^{l}\rangle -\langle j^{l}\rangle _{0}$, versus
the ratio $L/z_{1}$ for $\tilde{\alpha}=\pi /2$, $ma=1$, $z_{2}/z_{1}=2$, $%
z/z_{1}=1.5$.

\begin{figure}[tbph]
\begin{center}
\epsfig{figure=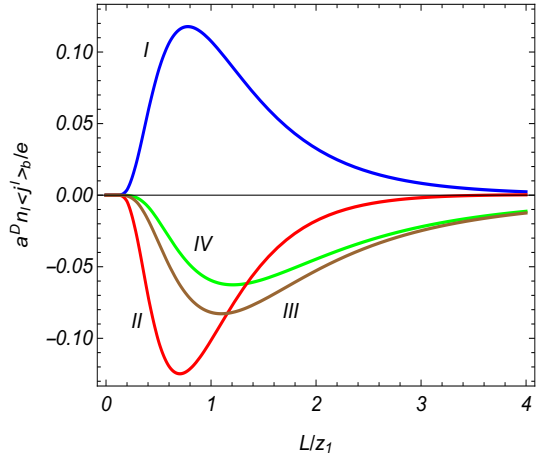,width=7.cm,height=6.cm}
\end{center}
\caption{The brane-induced contribution in the current density as a function
of the length of the compact dimension. The graphs are plotted for $\tilde{%
\protect\alpha}=\protect\pi /2$, $ma=1$, $z_{2}/z_{1}=2$, $z/z_{1}=1.5$.}
\label{fig4}
\end{figure}

\section{Second class of boundary conditions}

\label{sec:jlBC2}

For the normal $n_{\mu }^{(j)}$ to the brane at $z=z_{j}$ and for the Dirac
matrices we have the relation $(i\gamma ^{\mu }n_{\mu }^{(j)})^{2}=1$. This
means that $i\gamma ^{\mu }n_{\mu }^{(j)}$ has eigenvalues $\pm 1$. If we
assume that on the brane $i\gamma ^{\mu }n_{\mu }^{(j)}\psi =\pm \psi $ then
for both the signs one gets $n_{\mu }^{(j)}j^{\mu }=0$ for $z=z_{j}$ and
quantum numbers are not lost through the brane. In the discussion above we
have considered the boundary condition with the lower sign. Equally well
acceptable boundary condition is the one with the upper sign (see also \cite%
{Berr87}):%
\begin{equation}
(1-i\gamma ^{\mu }n_{\mu }^{(j)})\psi =0,\;z=z_{j},  \label{BC2}
\end{equation}%
with $j=1,2$. The positive and negative energy mode functions for these
conditions are still presented in the form (\ref{psiPM}). The boundary
conditions on the left and right branes are reduced to the equations $%
Z_{ma-1/2}(\lambda z_{1})=0$ and$\;Z_{ma+1/2}(\lambda z_{2})=0$,
respectively. Now the eigenvalues of the quantum number $\lambda $ are
determined from the equation
\begin{equation}
g_{\nu -1,\nu }(\lambda z_{1},\lambda z_{2})=0.  \label{eig2}
\end{equation}%
The corresponding positive roots with respect to $\lambda z_{1}$ will be
denoted by $\lambda _{n}^{(-)}=\lambda z_{1}$. For large values of $\lambda $%
, by using the asymptotic expressions for the Bessel functions, to the
leading order the equation (\ref{eig2}) is reduced to $\cos \left[ \lambda
\left( z_{2}-z_{1}\right) \right] =0$. For a massless field this equation is
exact. Hence, for large values of $n$ one has the asymptotic expression $%
\lambda _{n}^{(-)}\approx \pi (n-1/2)/\left( z_{2}/z_{1}-1\right) $. In the
Minkowski bulk two problems with boundary conditions (\ref{Bagbc}) and (\ref%
{BC2}) differ by rearrangement of two planar boundaries and the VEVs in the
region between the plates are the same. For the AdS bulk the boundaries have
nonzero extrinsic curvature and that is not the case.

In a way similar to that we have described for the boundary conditions (\ref%
{Bagbc}), for the mode functions one gets%
\begin{eqnarray}
\psi _{\beta }^{(+)} &=&B_{\beta }^{(+)}z^{\frac{D+1}{2}}e^{i\mathbf{kx}%
-i\omega t}\left(
\begin{array}{c}
\frac{\mathbf{k\chi }\chi _{0}^{\dagger }+i\lambda -\omega }{\omega }%
g_{ma-1/2,ma+s/2}(\lambda z_{1},\lambda z)w^{(\sigma )} \\
i\chi _{0}^{\dagger }\frac{\mathbf{k\chi }\chi _{0}^{\dagger }+i\lambda
+\omega }{\omega }g_{ma-1/2,ma-s/2}(\lambda z_{1},\lambda z)w^{(\sigma )}%
\end{array}%
\right) ,  \notag \\
\psi _{\beta }^{(-)} &=&B_{\beta }^{(-)}z^{\frac{D+1}{2}}e^{i\mathbf{kx}%
+i\omega t}\left(
\begin{array}{c}
i\chi _{0}\frac{\mathbf{k\chi }^{\dagger }\chi _{0}-i\lambda +\omega }{%
\omega }g_{ma-1/2,ma+s/2}(\lambda z_{1},\lambda z)w^{(\sigma )} \\
\frac{\mathbf{k\chi }^{\dagger }\chi _{0}-i\lambda -\omega }{\omega }%
g_{ma-1/2,ma-s/2}(\lambda z_{1},\lambda z)w^{(\sigma )}%
\end{array}%
\right) ,  \label{psiPM22}
\end{eqnarray}%
with the normalization coefficients
\begin{equation}
\left\vert B_{\beta }^{(\pm )}\right\vert ^{2}=\frac{(2\pi )^{2-p}\lambda }{%
32V_{q}a^{D}z_{1}}T_{ma-1/2}^{(-)}(\eta ,\lambda _{n}^{(-)}),  \label{Bp}
\end{equation}%
where%
\begin{equation}
T_{\nu }^{(-)}(\eta ,x)=x\left[ \frac{J_{ma-1/2}^{2}(x)}{J_{ma+1/2}^{2}(\eta
x)}-1\right] ^{-1}.  \label{Tnum}
\end{equation}

Likewise in the previous case the charge density and the components of the
current density along uncompact dimensions vanish. The components along
compact dimensions do not depend on the value of the parameter $s$ and are
given by
\begin{eqnarray}
\langle j^{l}\rangle &=&-\frac{\pi ^{2}\left( 4\pi \right) ^{-p/2}eNz^{D+2}}{%
4\Gamma (p/2)V_{q}a^{D+1}z_{1}}\sum_{\mathbf{n}_{q}}k_{l}\int_{0}^{\infty
}dk_{(p)}\,k_{(p)}^{p-1}\sum_{n=1}^{\infty }\frac{\lambda
_{n}^{(-)}T_{ma-1/2}^{(-)}(\eta ,\lambda _{n}^{(-)})}{\sqrt{\lambda
_{n}^{(-)2}+z_{1}^{2}k^{2}}}  \notag \\
&&\times \sum_{j=\pm 1}g_{ma-1/2,ma+j/2}^{2}(\lambda _{n}^{(-)},\lambda
_{n}^{(-)}z/z_{1}).  \label{jlm}
\end{eqnarray}%
For a massless field the equation for the eigenvalues $\lambda _{n}^{(-)}$
coincides with that for $\lambda _{n}$ in the case of the bag boundary
condition and the current densities coincide as well. The summation over the
eigenmodes $\lambda _{n}^{(-)}$ can be done by using the formula (\ref%
{SumFormAP}) with $\mu =ma+1/2$ and $\delta =-1$. The part with the first
term in the right-hand side of (\ref{SumFormAP}) gives the contribution to
the VEV from the left brane when the right one is absent. It is given by the
expression%
\begin{equation}
\langle j^{l}\rangle ^{(1)}=\langle j^{l}\rangle ^{(0)}-\frac{A_{p}eNz^{D+2}%
}{V_{q}a^{D+1}}\sum_{\mathbf{n}_{q}}k_{l}\int_{k_{(q)}}^{\infty
}du\,u(u^{2}-k_{(q)}^{2})^{\frac{p-1}{2}}\frac{I_{ma-1/2}(z_{1}u)}{%
K_{ma-1/2}(z_{1}u)}\sum_{j=\pm 1}jK_{ma+j/2}^{2}(zu).  \label{jl1b2}
\end{equation}%
The current density in the region between the branes is presented in the form%
\begin{eqnarray}
\langle j^{l}\rangle &=&\langle j^{l}\rangle ^{(0)}+\frac{A_{p}eNz^{D+2}}{%
V_{q}a^{D+1}}\sum_{\mathbf{n}_{q}}k_{l}\int_{k_{(q)}}^{\infty
}du\,u(u^{2}-k_{(q)}^{2})^{\frac{p-1}{2}}\,\left[ \frac{K_{\nu
-1}(z_{1}u)I_{\nu }(z_{2}u)}{I_{\nu -1}(z_{1}u)K_{\nu }(z_{2}u)}+1\right]
^{-1}  \notag \\
&&\times \sum_{j=\pm 1}j\left[ \frac{K_{\nu -1}(z_{1}u)}{I_{\nu -1}(z_{1}u)}%
I_{ma+j/2}^{2}(zu)-\frac{I_{\nu }(z_{2}u)}{K_{\nu }(z_{2}u)}%
K_{ma+j/2}^{2}(zu)\right.  \notag \\
&&\left. +2jI_{ma+j/2}(zu)K_{ma+j/2}(zu)\right] .  \label{jl2BC}
\end{eqnarray}%
An equivalent representations, similar to (\ref{jl4}), (\ref{jl6}) and (\ref%
{jl7}), can also be obtained for the boundary condition (\ref{BC2}). For the
integrated current one gets the formula that is obtained from (\ref{jlint})
with the replacement $\lambda _{n}\rightarrow \lambda _{n}^{(-)}$ and the
relation (\ref{relint}) remains the same.

Similar to (\ref{jlLargez2}), in the limit $z_{2}\rightarrow \infty $ the
contribution of the second brane in the VEV\ of the current density is
suppressed by the factor $e^{-2z_{2}k_{(q)}^{(0)}}$. For the limit $%
z_{1}\rightarrow 0$ two cases should be considered separately. In the case $%
ma>1/2$ one gets%
\begin{eqnarray}
\langle j^{l}\rangle &\approx &\langle j^{l}\rangle ^{(2)}-\frac{%
2A_{p}eN(z_{1}/2)^{2ma-1}z^{D+2}}{V_{q}a^{D+1}\Gamma \left( \nu -1\right)
\Gamma \left( \nu \right) }\sum_{\mathbf{n}_{q}}k_{l}\,  \notag \\
&&\times \int_{k_{(q)}}^{\infty }du\,u^{2ma}(u^{2}-k_{(q)}^{2})^{\frac{p-1}{2%
}}\sum_{j=\pm 1}j\frac{G_{\nu ,ma+j/2}^{2}(z_{2}u,zu)}{I_{\nu }^{2}(z_{2}u)},
\label{jlb2smz1}
\end{eqnarray}%
where%
\begin{equation}
\langle j^{l}\rangle ^{(2)}=\langle j^{l}\rangle ^{(0)}+\frac{A_{p}eNz^{D+2}%
}{V_{q}a^{D+1}}\sum_{\mathbf{n}_{q}}k_{l}\int_{k_{(q)}}^{\infty
}du\,u(u^{2}-k_{(q)}^{2})^{\frac{p-1}{2}}\sum_{j=\pm 1}j\frac{K_{\nu
}(z_{2}u)}{I_{\nu }(z_{2}u)}I_{ma+j/2}^{2}(zu).  \label{jlbb2}
\end{equation}%
In this case the contribution induced by the first brane tends to zero as $%
z_{1}^{2ma-1}$. For $ma=1/2$ the corresponding decay is logarithmic, as $%
1/\ln (z_{1}/z_{2})$. For $ma<1/2$, to the leading order one gets%
\begin{eqnarray}
\langle j^{l}\rangle &\approx &\langle j^{l}\rangle ^{(2)}-\frac{%
2A_{p}eNz^{D+2}}{\pi V_{q}a^{D+1}}\cos \left( ma\pi \right) \sum_{\mathbf{n}%
_{q}}k_{l}  \notag \\
&&\times \,\int_{k_{(q)}}^{\infty }du\,u\frac{(u^{2}-k_{(q)}^{2})^{\frac{p-1%
}{2}}}{I_{\nu }(z_{2}u)I_{-\nu }(z_{2}u)}\sum_{j=\pm 1}jG_{\nu
,ma+j/2}^{2}(z_{2}u,zu).  \label{jlb2smz1b}
\end{eqnarray}%
This leading term is different from that in the first case (see (\ref%
{jlb2smz1})). Note that if we consider a problem with the boundary condition
(\ref{BC2}) on the right brane but with the condition (\ref{Bagbc}) on the
left one, the limiting transition $z_{1}\rightarrow 0$ in the range of the
field mass $ma<1/2$ is completely different, the left brane induced
contribution behaves as $z_{1}^{2ma+1}$. The last term in (\ref{jlb2smz1b})
for the boundary condition (\ref{BC2}), in some sense, can be considered as
a memory from the left brane when its location tends to the AdS boundary.
This kind of memory is absent for the condition (\ref{Bagbc}).

In figures \ref{fig1}-\ref{fig4}, the curves corresponding to the boundary
condition (\ref{BC2}) are designated by II. As it has been mentioned above,
in the case of a massless field the current densities for the boundary
conditions (\ref{Bagbc}) and (\ref{BC2}) coincide. That is seen from figure %
\ref{fig2}. For a massive field the brane-induced contributions to the
current density for the boundary condition (\ref{BC2}) can be essentially
larger when compared with the brane-free part and the brane-induced part in
the case of the condition (\ref{Bagbc}).

\section{Currents in $Z_{2}$-symmetric models with two branes}

\label{sec:Brane}

With the results given above we can investigate the current density in
higher dimensional generalizations of Randall-Sundrum type braneworlds \cite%
{Rand99} with two branes and with a compact subspace. In these models the
coordinate $y$ is compactified on an orbifold $S^{1}/Z_{2}$ of length $b$,
with $-b\leqslant y\leqslant b$. The branes are located at the points $y=0$
and $y=b$ and the line element is given by (\ref{ds2}) where the warp factor
$e^{-2y/a}$ must be replaced by $e^{-2|y|/a}$. The original Randall-Sundrum
model has a single extra dimension, corresponding to $D=4$, and only the
gravitational field propagates on the bulk. However, in braneworld models
motivated from string theories we expect the presence of extra compact
dimensions and also extra bulk fields. Here we consider more general setup
with the locations of the branes at $y=y_{j}$, $j=1,2$.

In braneworl models, the boundary conditions on the branes for bulk fields
are obtained from the $Z_{2}$-symmetry. For a brane at $y=y_{j}$ and for a
fermionic field $\psi (x)$ one has $\psi (x^{i},y_{j}-y)=M_{j}\psi
(x^{i},y-y_{j})$, where $M_{j}$ is a $N\times N$ matrix. From the invariance
of the fermionic action under the $Z_{2}$ identification it can be seen that
(see \cite{Flac01b,Bell18}) this matrix have the form%
\begin{equation}
M_{j}=-u_{j}s\,\mathrm{diag}(1,-1),  \label{M}
\end{equation}%
where $u_{j}=\pm 1$ and we have extracted the factor $s$ for convenience.
With this transformation matrix, the boundary condition for the modes (\ref%
{psiPM}) on the brane $y=y_{j}$ is reduced to $Z_{ma+u_{j}/2}(\lambda
z_{j})=0$ for both the positive and negative energy solutions. We also see
that with the choice (\ref{M}) (opposite signs of the matrix $M$ for $s=1$
and $s=-1$), the boundary condition is the same for $s=\pm 1$. As a result,
in the geometry of two branes one has four different combinations of the
boundary conditions corresponding to different choices of $u_{j}$ in the set
$(u_{1},u_{2})$ (for different combinations of boundary conditions imposed
on fermionic fields in two-brane models see also \cite{Chan05}).

For given $(u_{1},u_{2})$, the mode functions obeying the boundary condition
on the brane $y=y_{1}$ are presented as%
\begin{eqnarray}
\psi _{\beta }^{(+)}(x) &=&D_{\beta }^{(+)}z^{\frac{D+1}{2}}e^{i\mathbf{kx}%
-i\omega t}\left(
\begin{array}{c}
\frac{\mathbf{k\chi }\chi _{0}^{\dagger }+i\lambda -\omega }{\omega }%
g_{ma+u_{1}/2,ma+s/2}(\lambda z_{1},\lambda z)w^{(\sigma )} \\
i\chi _{0}^{\dagger }\frac{\mathbf{k\chi }\chi _{0}^{\dagger }+i\lambda
+\omega }{\omega }g_{ma+u_{1}/2,ma-s/2}(\lambda z_{1},\lambda z)w^{(\sigma )}%
\end{array}%
\right) ,  \notag \\
\psi _{\beta }^{(-)}(x) &=&D_{\beta }^{(-)}z^{\frac{D+1}{2}}e^{i\mathbf{kx}%
+i\omega t}\left(
\begin{array}{c}
i\chi _{0}\frac{\mathbf{k\chi }^{\dagger }\chi _{0}-i\lambda +\omega }{%
\omega }g_{ma+u_{1}/2,ma+s/2}(\lambda z_{1},\lambda z)w^{(\sigma )} \\
\frac{\mathbf{k\chi }^{\dagger }\chi _{0}-i\lambda -\omega }{\omega }%
g_{ma+u_{1}/2,ma-s/2}(\lambda z_{1},\lambda z)w^{(\sigma )}%
\end{array}%
\right) .  \label{psiPMbr}
\end{eqnarray}%
From the boundary condition on the brane $y=y_{2}$ it follows that now the
eigenvalues for $\lambda $ are roots of the equation%
\begin{equation}
g_{ma+u_{1}/2,ma+u_{2}/2}(\lambda z_{1},\lambda z_{2})=0.  \label{Eigbr}
\end{equation}%
For the normalization coefficients we get%
\begin{equation}
\left\vert D_{\beta }^{(\pm )}\right\vert ^{2}=\frac{\lambda
T_{ma+u_{1}/2}^{(u_{1},u_{2})}(z_{2}/z_{1},\lambda z_{1})}{32N_{0}z_{1}(2\pi
)^{p-2}V_{q}a^{D}}.  \label{Cbr}
\end{equation}%
where%
\begin{equation}
T_{ma+u_{1}/2}^{(u_{1},u_{2})}(\eta ,x)=x\left[ \frac{J_{ma+u_{1}/2}^{2}(x)}{%
J_{ma+u_{2}/2}^{2}(\eta x)}-1\right] ^{-1}.  \label{Tbr}
\end{equation}%
Note that the equation (\ref{Eigbr}) corresponds to the boundary conditions
\begin{equation}
(1-(-1)^{j}iu_{j}\gamma ^{\mu }n_{\mu }^{(j)})\psi (x)=0,\;z=z_{j},
\label{BC3}
\end{equation}%
on the branes.

In $Z_{2}$-symmetric braneworld models the normalization integral goes over
the two copies of the region $y_{1}\leqslant y\leqslant y_{2}$ and in (\ref%
{Cbr}) $N_{0}=2$. In the analog of the problem we have considered in the
previous sections with two branes $y=y_{j}$ and with the boundary conditions
$Z_{ma+u_{j}/2}(\lambda z_{j})=0$ on them, in the region $y_{1}\leqslant
y\leqslant y_{2}$ one should take $N_{0}=1$ in (\ref{Cbr}). Note that for $%
u_{1}=u_{2}$ and for large values of $\lambda $ the equation (\ref{Eigbr})
is reduced to $\sin \left[ \lambda \left( z_{2}-z_{1}\right) \right] =0$ and
for the corresponding modes one has asymptotic expression $\lambda
z_{1}=\lambda _{ma+u_{1}/2,n}^{(0)}\approx \pi n/\left( z_{2}/z_{1}-1\right)
$ with large $n$. For massless fields this expression is exact.

Now we see that the current densities in $Z_{2}$-symmetric braneworlds with
the combination of the boundary conditions on the branes corresponding to $%
(u_{1},u_{2})=(+1,-1)$ are obtained from the results in Section \ref%
{sec:Curr} with an additional coefficient 1/2. For the set of boundary
conditions with $(u_{1},u_{2})=(-1,+1)$ the corresponding current density is
obtained from the formulas in Section \ref{sec:jlBC2} (again, with the
factor 1/2). The current densities for the combinations of the boundary
conditions corresponding to $(u_{1},u_{2})=(+1,+1)$ and $%
(u_{1},u_{2})=(-1,-1)$ can be considered in a similar way we have described
in Section \ref{sec:Curr} for the case $(u_{1},u_{2})=(+1,-1)$. The VEV of
the current density along the $l$th compact dimension is presented in the
form similar to (\ref{jl1}), where now $\lambda $ is the root of the
equation (\ref{Eigbr}) with $u_{2}=u_{1}$. The summation formula for the
series over these roots is obtained from (\ref{SumFormAP}) with $\delta =0$
and $\mu =ma+u_{1}/2$ and the further transformation for the VEV is similar
to that in Section \ref{sec:Curr}. The final expression for the current
density in the region between the branes takes the form%
\begin{eqnarray}
\langle j^{l}\rangle &=&\langle j^{l}\rangle ^{(0)}+\frac{u_{1}A_{p}eNz^{D+2}%
}{N_{0}V_{q}a^{D+1}}\sum_{\mathbf{n}_{q}}k_{l}\int_{k_{(q)}}^{\infty
}du\,u(u^{2}-k_{(q)}^{2})^{\frac{p-1}{2}}\,\left[ \frac{K_{\mu
}(z_{1}u)I_{\mu }(z_{2}u)}{K_{\mu }(z_{2}u)I_{\mu }(z_{1}u)}-1\right] ^{-1}
\notag \\
&&\times \sum_{j=\pm 1}j\left[ \frac{I_{\mu }(z_{2}u)}{K_{\mu }(z_{2}u)}%
K_{ma+j/2}^{2}(zu)+\frac{K_{\mu }(z_{1}u)}{I_{\mu }(z_{1}u)}%
I_{ma+j/2}^{2}(zu)\right.  \notag \\
&&\left. -2ju_{1}I_{ma+j/2}(zu)K_{ma+j/2}(zu)\right] ,  \label{jlZ2}
\end{eqnarray}%
where $\mu =ma+u_{1}/2$ and $u_{1}=\pm 1$. The single brane contribution to
the vacuum current density for the brane at $y=y_{1}$ is given by (\ref{jlRb}%
) for $u_{1}=1$ and by the last term in (\ref{jl1b2}) for $u_{1}=-1$ (with
additional factors $1/N_{0}$ for $Z_{2}$-symmetric braneworlds). We can also
obtain an alternative representation similar to (\ref{jl7}).

In figures \ref{fig1}-\ref{fig4}, the graphs for the current densities (\ref%
{jlZ2}) (with $N_{0}=1$) in the cases $u_{1}=u_{2}=+1$ and $u_{1}=u_{2}=-1$
are designated by roman numerals III and IV, respectively. Note that one has
the relation $g_{-\mu ,-\mu }(x,u)=g_{\mu ,\mu }(x,u)$ and the eigenmodes
for $\lambda $ in these cases coincide for a massless field. From here it
follows that the current densities corresponding to III and IV are the same
in the limit $m\rightarrow 0$. This is seen from figure \ref{fig2}.

In the Randall-Sundrum scenario the standard model fields are localized on
the brane $z=z_{2}$ (visible or infrared brane). The current density on that
brane is a source of magnetic fields having components in the uncompact
subspace as well. It is of interest to separate the parts in the current
density on the visible brane induced by the presence of the hidden (or
ultraviolet) brane with the location $z=z_{1}$. By using the expressions
given above, we can combine the hidden brane-induced contributions for
different combinations of the boundary conditions, specified by the set $%
(u_{1},u_{2})$, in a single expression
\begin{eqnarray}
\left[ \langle j^{l}\rangle -\langle j^{l}\rangle ^{(2)}\right] _{z=z_{2}}
&=&\frac{A_{p}eNz_{2}^{D}}{N_{0}V_{q}a^{D+1}}\sum_{\mathbf{n}%
_{q}}k_{l}\int_{k_{(q)}}^{\infty }du\,\frac{I_{ma+u_{1}/2}(z_{1}u)}{%
uI_{ma+u_{2}/2}(z_{2}u)}  \notag \\
&&\times \,\frac{(u^{2}-k_{(q)}^{2})^{\frac{p-1}{2}}}{%
G_{ma+u_{1}/2,ma+u_{2}/2}(z_{1}u,z_{2}u)}.  \label{jlhid}
\end{eqnarray}%
In braneworld models of the Randall-Sundrum type, in order to solve the
hierarchy problem between the Planck and electroweak energy scales, it is
assumed that $(y_{2}-y_{1})\gg a$. Under this condition one has $%
z_{2}/z_{1}\gg 1$ and the asymptotic behavior of (\ref{jlhid}) depends on
the lengths of compact dimensions. For $z_{1}/L_{i}\gtrsim 1$, in the
integration range of (\ref{jlhid}) one has $z_{2}x/z_{1}\gg 1$. In a way
similar to that we have used for (\ref{jlLargez2}), it can be seen that to
the leading order one has%
\begin{equation}
\left[ \langle j^{l}\rangle -\langle j^{l}\rangle ^{(2)}\right]
_{z=z_{2}}\approx -\frac{eNz_{2}^{q+2}\tilde{\alpha}_{l}\left(
z_{2}k_{(q)}^{(0)}\right) ^{(p-1)/2}}{2^{p+1}\pi
^{(p-1)/2}N_{0}V_{q}L_{l}a^{D+1}}\frac{I_{ma+u_{1}/2}(z_{1}k_{(q)}^{(0)})}{%
K_{ma+u_{1}/2}(z_{1}k_{(q)}^{(0)})}e^{-2z_{2}k_{(q)}^{(0)}},  \label{jlhidas}
\end{equation}%
where $k_{(q)}^{(0)}$ is defined by (\ref{kq0}). For $z_{2}/L_{i}\lesssim 1$
and $z_{2}/z_{1}\gg 1$ the asymptotic expression of (\ref{jlhid}) is found
in a way similar to that used above for the limit $z_{1}\rightarrow 0$. For $%
ma+u_{1}/2>0$ we can see that the hidden brane contribution in the current
density on the visible brane behaves like $(z_{1}/z_{2})^{2ma+u_{1}}$. In
the case $ma+u_{1}/2<0$ (for nonnegative $m$ this implies $u_{1}=-1$) one
gets%
\begin{equation}
\left[ \langle j^{l}\rangle -\langle j^{l}\rangle ^{(2)}\right]
_{z=z_{2}}\approx \frac{2u_{2}A_{p}eNz_{2}^{q+2}}{\pi N_{0}V_{q}a^{D+1}}%
\sum_{\mathbf{n}_{q}}k_{l}\int_{z_{2}k_{(q)}}^{\infty }dx\,\,\frac{\cos
\left( \pi ma\right) (x^{2}-z_{2}k_{(q)}^{2})^{\frac{p-1}{2}}}{%
xI_{ma+u_{2}/2}(x)I_{-ma-u_{2}/2}(x)},  \label{jlhidas2}
\end{equation}%
and the leading term does not depend on $z_{1}$.

\section{P- and T-reversal symmetric odd-dimensional models and applications
to curved graphene tubes}

\label{sec:OddD}

In this section we consider features of fermionic models in odd-dimensional
spacetimes. As it has been already mentioned, for even $D$ there are two
inequivalent irreducible representations of the Clifford algebra. For flat
spacetime Dirac matrices $\gamma ^{(b)}$ with $b=0,\ldots ,D-1$, we
introduce the $2^{D/2}\times 2^{D/2}$ matrix $\gamma
=\prod_{b=0}^{D-1}\gamma ^{(b)}$. Then we can take the matrix $\gamma ^{(D)}$
in the form $\gamma _{(s)}^{(D)}=s\gamma $ for $D=4n$ and in the form $%
\gamma _{(s)}^{(D)}=si\gamma $ for $D=4n-2$, where $n=1,2,\ldots $. Here, $%
s=+1$ and $s=-1$ correspond to two irreducible representations of the
Clifford algebra. For the curved spacetime matrix $\gamma _{(s)}^{D}$,
corresponding to the geometry described by (\ref{ds2}), one can take $\gamma
_{(s)}^{D}=(z/a)\gamma _{(s)}^{(D)}$. Let us denote by $\psi _{(s)}$ the
fermioinc field realizing the representations with given $s$. The mass term
in the corresponding Lagrangian density $\mathcal{L}_{(s)}=\bar{\psi}%
_{(s)}[i\gamma _{(s)}^{\mu }(\partial _{\mu }+\Gamma _{\mu }^{(s)})-m]\psi
_{(s)}$, with the set of Dirac matrices $\gamma _{(s)}^{\mu }=(\gamma
^{0},\gamma ^{1},\cdots \gamma ^{D-1},\gamma _{(s)}^{D})$ and the related
spin connection $\Gamma _{\mu }^{(s)}$, is not invariant under the charge
conjugation ($C$) and parity transformation ($P$) in spatial dimensions $%
D=4n $, and under the $P$-transformation and the time reversal ($T$) in $%
D=4n+2$.

We can construct fermionic models in odd-dimensional spacetimes, invariant
under the $C$-, $P$- and $T$-transformations, combining two fields $\psi
_{(s)}$ with the Lagrangian density $\mathcal{L}=\sum_{s=\pm 1}\mathcal{L}%
_{(s)}$. By approproate transformations of the fields one can make this
combined Lagrangian density invariant under the $C$-, $P$- and $T$%
-transformations. Introducing $2N\times 2N$ matrices $\gamma ^{(2N)\mu }=%
\mathrm{diag}(\gamma _{(+1)}^{\mu },\gamma _{(-1)}^{\mu })$, with $N=2^{D/2}$%
, and the corresponding spin connection $\Gamma _{\mu }^{(2N)}$, we can
combine two fields $\psi _{(s)}$ in a single $2N$-component field $\Psi
=(\psi _{(+1)},\psi _{(-1)})^{T}$ with the Lagrangian density
\begin{equation}
\mathcal{L}=\bar{\Psi}[i\gamma ^{(2N)\mu }(\partial _{\mu }+\Gamma _{\mu
}^{(2N)})-m]\Psi  \label{L2N}
\end{equation}%
and the current density operator $J^{\mu }=e\bar{\Psi}\gamma ^{(2N)\mu }\Psi
$. An alternative representation of the model with two fields is obtained by
making the field transformations $\psi _{(+1)}^{\prime }=\psi _{(+1)}$, $%
\psi _{(-1)}^{\prime }=\gamma \psi _{(-1)}$. The combined Lagrangian density
is presented as $\mathcal{L}=\sum_{s=\pm 1}\bar{\psi}_{(s)}^{\prime
}[i\gamma ^{\mu }\left( \partial _{\mu }+\Gamma _{\mu }\right) -sm]\psi
_{(s)}^{\prime }$, where $\Gamma _{\mu }$ is the spin connection for the set
of Dirac matrices $\gamma ^{\mu }=\gamma _{(+1)}^{\mu }$. In this
representation the Lagrangian densities for the fields with $s=+1$ and $s=-1$
differ by the sign of the mass term.

In the system of two fermionic fields $\psi _{(s)}$ the VEV of the current
density is the sum of the VEVs coming from the separate fields $\langle
J^{\mu }\rangle =\sum_{s=\pm 1}\langle j_{(s)}^{\mu }\rangle $. As we have
seen above, if the boundary and periodicity conditions for the fields $\psi
_{(s)}$ are the same, then the separate contributions $\langle j_{(s)}^{\mu
}\rangle $ are the same as well and the total current density is obtained
from the expressions given above with an additional factor 2. However, both
the boundary conditions and the phases in the periodicity conditions can be
different for $s=+1$ and $s=-1$. In particular, we can combine various
boundary conditions of the form (\ref{BC3}) with different values of the
parameters $u_{j}$ for separate fields. The corresponding VEVs for the
current density for $s=+1$ and $s=-1$ are obtained from the formulas given
above. An example of a condensed matter realization of the problem with
different phases in the periodicity conditions for the fields $\psi _{(+1)}$
and $\psi _{(-1)}$ is provided by semiconducting carbon nanotubes (see
below).

Among the most important applications of $D=2$ fermionic models are the so
called Dirac materials. They include graphene, topological insulators and
Weyl semimetals. For these materials the long-wavelength excitations of the
electronic subsystem are well described by the Dirac equation with the
velocity of light replaced by the Fermi velocity $v_{F}$. Here we specify
the consideration for graphene. For a given quantum number $S=\pm 1$,
corresponding to spin degrees of freedom, the analog of the Lagrangian
density (\ref{L2N}) with $N=2$ is written for a 4-component spinor field $%
\Psi _{S}=(\psi _{+,AS},\psi _{+,BS},\psi _{-,AS},\psi _{-,BS})^{T}$. Here,
the indices $+$ and $-$ correspond to two inequivalent Fermi points at the
corners of the Brillouin zone (points $\mathbf{K}_{+}$ and $\mathbf{K}_{-}$)
and the indices $A$ and $B$ correspond to the triangular sublattices of the
graphene hexagonal lattice. The separate components of $\Psi _{S}$ present
the corresponding amplitude of the electron wave function (see, for example,
\cite{Gusy07}). For the fields we have introduced before one has $\psi
_{(\pm 1)}=\left( \psi _{\pm ,AS},\psi _{\pm ,BS}\right) ^{T}$. The mass
term in the Dirac equation is expressed in terms of the energy gap $\Delta $
by the relation $m=\Delta /v_{F}^{2}$. This gap can be generated by a number
of mechanisms. For the corresponding Compton wavelength one has $a_{C}=\hbar
v_{F}/\Delta $.

The graphene is an interesting arena for investigation of various kinds of
topological effects in field theory (for topological effects in condensed
matter physics see, for example, \cite{Gupt18}). The graphene made
structures with nontrivial topology include fullerens, carbon nanotubes and
nanoloops, and graphitic cones. They all have been experimentally observed.
The spatial topology of the problem with $D=2$, we have considered above,
corresponds to that for carbon nanotubes (topology $S^{1}\times R^{1}$). In
graphene nanotubes the phases in the periodicity conditions (\ref{PCo}) for
the fields $\psi _{(s)}$ depend on the chirality of the tube. For metallic
nanotubes one has $\alpha _{1}\equiv \alpha =0$ for both the fields $s=+1$
and $s=-1$. For semiconducting nanotubes the phases have opposite signs for
spinors corresponding to the points $\mathbf{K}_{\pm }$ and $\alpha =\pm
2\pi /3$.

\begin{figure}[tbph]
\begin{center}
\epsfig{figure=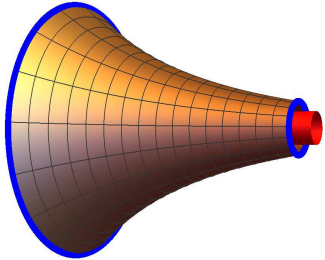,width=5.5cm,height=5.5cm}
\end{center}
\caption{The $D=2$ spatial geometry with two edges embedded in $R^{3}$. }
\label{fig5}
\end{figure}

For a cylindrical nanotube rolled-up from a planar graphene sheet the
spacetime geometry is flat. The corresponding VEV of the fermionic current
density induced by the threading magnetic flux has been discussed in \cite%
{Bell10} for infinite length tubes and in \cite{Bell13} for finite length
tubes. For the problem under consideration in the present paper, the spatial
geometry, written in terms of the angular coordinate $\varphi =2\pi x^{1}/L$%
, $0\leqslant \varphi \leqslant 2\pi $, is given by the line element $%
dl^{2}=dy^{2}+(L/2\pi )^{2}e^{-2y/a}d\varphi ^{2}$ with $y_{1}\leqslant
y\leqslant y_{2}$. This describes a finite length curved circular tube with
the radius $r=Le^{-y/a}/2\pi $ depending on the coordinate along the tube
axis (for the curvature effects in graphene structures see also \cite%
{Kole09,Iori14}). The corresponding 2-dimensional surface with two edges,
embedded in 3-dimensional Euclidean space, is depicted in figure \ref{fig5}.
In the figure we have also shown the magnetic flux enclosed by the curved
tube. The graphene tubes with spatial geometry described by the line element
$dl^{2}$ have been discussed in \cite{Iori14}. The geometry corresponds to
Beltrami pseudosphere with Gaussian curvature $-1/a^{2}$. The generation of
a pseudosphere configurations from a planar graphene sheet has been recently
discussed in \cite{Morr19} (see also the references therein). The
corresponding curvature radius varies in the range $1.5\,\mathrm{nm}<a<74\,%
\mathrm{nm}$. Examples of wormhole geometries realized by curved graphene
sheets have been considered in \cite{Gonz10}. An important difference in the
geometry we consider is that $g_{00}=e^{-2y/a}\neq 1$. A number of
mechanisms have been discussed recently for generation of the nontrivial $%
g_{00}$-component of the metric tensor for the low-energy effective field
theory describing the dynamics of electrons in graphene. This can be done by
various types of external fields, by deformations of graphene lattice
(strains), and by the local variations in the Fermi velocity (for reviews
see \cite{Volo15}).

In graphene tubes with the geometry under consideration the current density
for a given spin $S$ is obtained by summing the contributions $\langle
j_{(s)}^{\mu }\rangle $ coming from the fields $\psi _{(s)}$ corresponding
to the points $\mathbf{K}_{\pm }$. In the expression for the corresponding
operator for spatial components an additional factor $v_{F}$ should be
added, $j_{(s)}^{\mu }=ev_{F}\bar{\psi}\gamma _{(s)}^{\mu }\psi $ (with $%
e=-|e|$ for electrons). The expressions for $\langle j_{(s)}^{\mu }\rangle $
are obtained from the formulas given above taking $D=2$, $p=0$, $q=1$. We
can also express the product $ma$ in terms of the Compton wavelength
corresponding to the energy gap as $ma=a/a_{C}$. In the absence of the
magnetic flux, the VEV of the current density vanishes in both metallic and
semiconducting graphene tubes. In metallic tubes the separate contributions $%
\langle j_{(s)}^{\mu }\rangle $ are zero, whereas in semiconducting tubes $%
\langle j_{(-1)}^{\mu }\rangle =-\langle j_{(+1)}^{\mu }\rangle $ (assuming
that the boundary conditions on the edges of the tube are the same for
separate fields) because of the opposite signs of the phases in the
periodicity conditions. Nonzero net currents may appear in the presence of
the magnetic flux enclosed by the tube. In the absence of the magnetic flux,
nonzero ground state currents in semiconducting tubes can be alternatively
generated by imposing different boundary conditions on the edges for
separate fields $\psi _{(+1)}$ and $\psi _{(-1)}$.

\begin{figure}[tbph]
\begin{center}
\begin{tabular}{cc}
\epsfig{figure=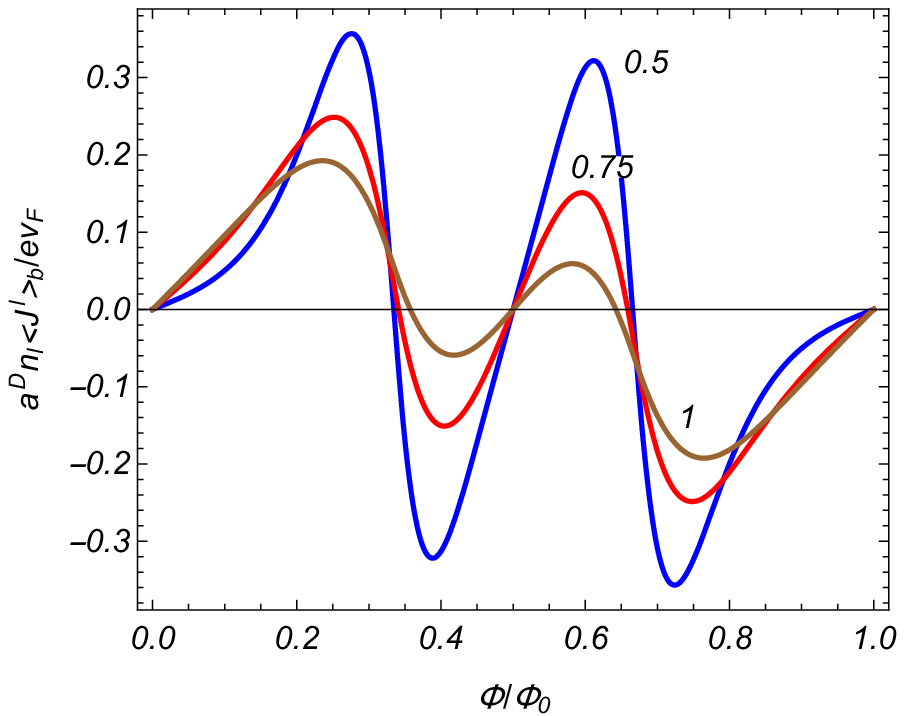,width=7.cm,height=5.5cm} & \quad %
\epsfig{figure=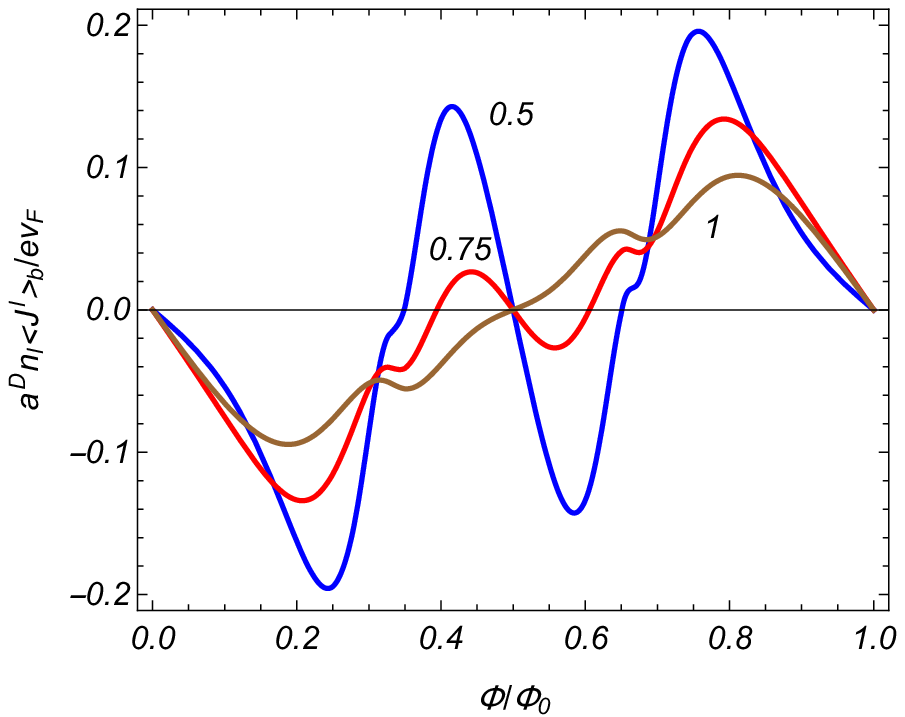,width=7.cm,height=5.5cm}%
\end{tabular}%
\end{center}
\caption{The edge-induced current density as a function of the magnetic flux
for semiconducting nanotubes. The left and right panels correspond to the
boundary conditions (\protect\ref{Bagbc}) and (\protect\ref{BC2}). The
graphs are plotted for $a/a_{C}=1$, $z_{2}/z_{1}=2$, $z/z_{1}=1.5$ and the
numbers near the curves are the values of the ratio $L/z_{1}$.}
\label{fig6}
\end{figure}

In figure \ref{fig6} we have plotted the edge contribution in the fermionic
current density, $\langle J^{\mu }\rangle _{b}=\langle J^{\mu }\rangle
-\langle J^{\mu }\rangle _{0}$, in semiconducting tubes as a function of the
enclosed magnetic flux (in units of flux quantum). The total current density
(for a given $S$) is obtained summing the current densities for the fields $%
\psi _{(+1)}$ and $\psi _{(-1)}$ with the phases in the periodicity
condition $2\pi /3$ and $-2\pi /3$, respectively. The left and right panels
correspond to the boundary conditions (\ref{Bagbc}) and (\ref{BC2}),
respectively. The graphs are plotted for $L/z_{1}=0.5,0.75,1$ (the numbers
near the curves) and for fixed $a/a_{C}=1$, $z_{2}/z_{1}=2$, $z/z_{1}=1.5$.

\begin{figure}[tbph]
\begin{center}
\epsfig{figure=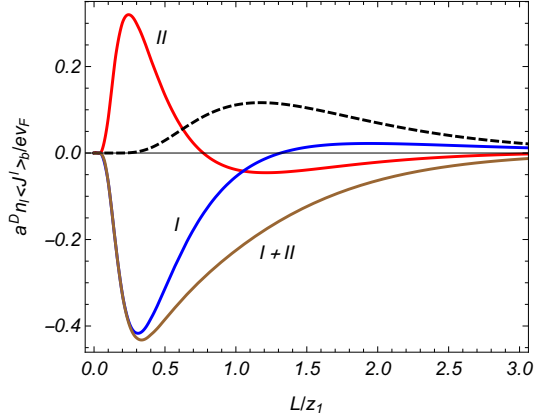,width=7.cm,height=5.5cm}
\end{center}
\caption{The edge-induced current density as a function of the tube
coordinate circumference for curved semiconducting nanotubes. The graphs are
plotted for different choices of the boundary conditions on the tube edges.
For the values of the related parameters see the text.}
\label{fig7}
\end{figure}

In figure \ref{fig7} the edge-induced current density is displayed as a
function of the tube coordinate circumference for semiconducting nanotube.
The curves I and II correspond to the boundary conditions (\ref{Bagbc}) and (%
\ref{BC2}), respectively, for both the fields $\psi _{(+1)}$ and $\psi
_{(-1)}$. The curve I+II and the dashed curve correspond to the situation
when the boundary condition (\ref{Bagbc}) is imposed for the field $\psi
_{(+1)}$ and the boundary condition (\ref{BC2}) for the field $\psi _{(-1)}$%
. The graphs I, II, I+II are plotted for the magnetic flux $\Phi =0.4\Phi
_{0}$ and the dashed graph corresponds to $\Phi =0$. For the values of the
remaining parameters we have taken $a/a_{C}=1$, $z_{2}/z_{1}=2$, $%
z/z_{1}=1.5 $. The dashed curve in figure \ref{fig7} presents an example
where a nonzero current density is generated in the absence of magnetic flux
by imposing different boundary conditions on separate fields corresponding
to different Fermi points.

Note that we have considered a model where the only interaction of the
fermionic field is with background classical gravitational and
electromagnetic fields. The effects of geometry, topology and boundaries in
models with four-fermion interactions have been discussed in \cite{Inag97}.
The geometrical and topological aspects of electronic interactions in
graphene and related materials are reviewed in \cite{Cort12}.

\section{Conclusion}

\label{sec:Conc}

Among the most important local characteristics of the vacuum state for
charged fields is the VEV of the current density. We have studied the
effects of two parallel branes on the current density in locally AdS
spacetime with a part of spatial dimensions (in Poincar\'{e} coordinates)
compactified to a torus. Along compact dimensions quasiperiodicity
conditions were imposed with general values of the phases and the presence
of a constant gauge field is assumed. The influence of the latter on the
physical properties of the vacuum state is of Aharonov-Bohm type and is
related to the nontrivial topology of the background geometry. On the branes
we have considered several types of boundary conditions including the ones
arising in $Z_{2}$-symmetric braneworld models. In the region between the
branes, the eigenvalues of the radial quantum number are zeros of the
combinations of the Bessel and Neumann functions. The mode sum for the VEV
of the current density contains series over those eigenvalues. In order to
find an integral representation, convenient in numerical calculations, we
have used a variant of the generalized Abel-Plana formula that allowed to
extract explicitly the brane-induced contributions. For all the boundary
conditions discussed, the VEVs of the charge density and of the components
of the current density along uncompact dimensions vanish.

In the investigation of the current density along compact dimensions, first
we have considered the case of the bag boundary condition that is the most
frequently one used for confinement of fermionic fields. In the region
between the branes the $l$th component is presented as (\ref{jl5}) where the
brane-induced contribution is explicitly extracted. The vacuum currents in
the absence of the branes were investigated in \cite{Bell17} and here we
were mainly concerned about the brane-induced effects. We have also provided
representations, given by (\ref{jl2}) and (\ref{jl6}), with the separated
contribution of the second brane when one adds it to the configuration with
a single brane. The effects of the phases in the periodicity conditions and
of the gauge field are encoded in the parameters $\tilde{\alpha}_{i}$. All
the contributions to the $l$th component of the current density are odd
periodic functions of $\tilde{\alpha}_{l}$ and even periodic functions of $%
\tilde{\alpha}_{i}$, $i\neq l$, with the period $2\pi $. In terms of the
magnetic flux enclosed by the compact dimension, this correspond to the
periodicity with the period equal to the flux quantum. An alternative
representation of the current density, well adapted for the investigation of
the near-brane asymptotic, is given by (\ref{jl7}). Unlike to the initial
representation (\ref{jl1}), the series over the eigenvalues of the radial
quantum number is exponentially convergent. The new representation also
explicitly shows the finiteness of the current density on the branes. The
latter feature is in clear contrast to the on-brane behavior of the fermion
condensate and of the VEV of the energy-momentum tensor having surface
divergences. The current density, integrated over the region between the
branes, is connected to the on-brane values of the current density by a
simple relation (\ref{relint}).

The general expression for the current density is rather complicated and, in
order to clarify its behavior as a function of the parameters, we have
considered various asymptotic limits. First of all, in the limit of large
curvature radius the result is obtained for the geometry of two parallel
plates in a locally Minkowski spacetime with a toroidal subspace, previously
discussed in \cite{Bell13}. For a massless fermionic field, the problem
under consideration is conformally related to the corresponding problem in
locally Minkowski bulk and the current density is given by a simple
expression (\ref{jlm0}). In the limit when the right brane tends to the AdS
horizon, for fixed location of the left brane and of the observation point,
the corresponding contribution to the current density is exponentially
suppressed by the factor $e^{-2z_{2}k_{(q)}^{(0)}}$, with $k_{(q)}^{(0)}$
defined by (\ref{kq0}). When the location of the left brane tends to the AdS
boundary, the corresponding contribution to the vacuum current decays like $%
z_{1}^{2ma+1}$.

If the length of the $l$th compact dimension $L_{l}$ is much smaller than
the other length scales in the problem, including the difference $%
z_{2}-z_{1} $, the brane-induced contribution to the current density along
that direction is suppressed by the factor $\exp [-2(z_{2}-z_{1})|\tilde{%
\alpha}_{l}|/L_{l}]$ and the total current is dominated by the brane-free
part. For large values of $L_{l}$, the current density is dominated by the
mode with the lowest value $\lambda =\lambda _{1}/z_{1}$ of the radial
quantum number and the current density is suppressed by the factor $\exp
[-L_{l}\sqrt{\lambda _{1}^{2}/z_{1}^{2}+k_{(q-1)}^{(0)2}}]$. The behavior of
the $l$th component of the current density for small values of the length $%
L_{i}$, $i\neq l$, crucially depends whether the phase $\tilde{\alpha}_{i}$,
$|\tilde{\alpha}_{i}|<\pi $, is zero or not. For $\tilde{\alpha}_{i}=0$ the
dominant contribution comes from the zero mode along the $i$th dimensions
and, to the leading order, the current density $\langle j^{l}\rangle $ is
expressed in terms of the corresponding current density in $D$-dimensional
spacetime with excluded $i$th dimension. In the case $\tilde{\alpha}_{i}\neq
0$, the VEV $\langle j^{l}\rangle $ is suppressed by the factor $%
e^{-2(z_{2}-z_{1})|\tilde{\alpha}_{i}|/L_{i}}$.

The investigation of the current density for the boundary condition (\ref%
{BC2}) is done in a way similar to that in the case of the bag boundary
condition. The corresponding current density in the region between the
branes is decomposed as (\ref{jl2BC}). For the mass range $ma<1/2$, an
important difference when compared to the bag boundary conditions appears in
the limit when the left brane goes to the AdS boundary ($z_{1}\rightarrow 0$%
). An additional contribution survives (last term in (\ref{jlb2smz1b})) that
can be interpreted as some kind of memory from the boundary condition we
have imposed on the brane at $z=z_{1}$. Yet another two classes of boundary
conditions arise in $Z_{2}$-symmetric braneworld models. They correspond to
the sets $(u_{1},u_{2})$ with $u_{1}=u_{2}$ in the conditions (\ref{BC3}).
The corresponding current densities are given by (\ref{jlZ2}) with $\mu
=ma+u_{1}/2$. The memory effect in the limit $z_{1}\rightarrow 0$ is present
for the boundary condition with $u_{1}=-1$. Depending on the boundary
conditions imposed, the presence of the branes can either increase or
decrease the current density. In braneworld models of the Randall-Sundrum
type the observers are localized on the right brane and it is of interest to
investigate the effects of the hidden brane on the current density on the
visible brane. The part of the vacuum current induced by the hidden brane is
given by (\ref{jlhid}). For the solution of the hierarchy problem between
the electroweak and Planck energy scales it is required to have $%
z_{2}/z_{1}\gg 1$. In this limit the behavior of the hidden brane-induced
current essentially depends on the lengths of compact dimensions and is
different for $z_{1}/L_{i}\gtrsim 1$ and $z_{2}/L_{i}\lesssim 1$.

In odd-dimensional spacetimes, the models with massive fermionic fields
realizing irreducible representations of the Clifford algebra are not parity
and time-reversal invariant. Fermionic models with parity and time-reversal
symmetry are constructed combining two fields corresponding to inequivalent
representations. If the periodicity conditions along compact dimensions and
the boundary conditions on the branes are the same for separate fields, when
the current densities for those fields are the same as well and the
expressions for the total current density is obtained from those presented
with an additional factor two. However, both the periodicity and boundary
conditions can be different for fields realizing inequivalent
representations of the Clifford algebra. An example of $D=2$ fermionic
system with that type of situation is provided by semiconducting carbon
nanotubes, with the electronic subsystem described by the Dirac model. In
the corresponding setup the phases for separate fields have opposite signs
and, in the absence of the magnetic flux, the corresponding current
densities cancel each other if the boundary conditions for the fields are
the same. In the case of different boundary conditions on the tube edges for
separate fields, a nonzero current can be generated in the absence of
magnetic flux. Curved graphene structures provide an important laboratory
for the investigation of curvature and topological effects in quantum field
theory. The special case $D=2$ of our model presents an exactly solvable
problem of that kind.

\section*{Acknowledgments}

A.A.S. was supported by Viktor Ambartsumian Research Fellowship 2019-2020.
A.A.S. gratefully acknowledges the hospitality of the INFN, Laboratori
Nazionali di Frascati (Frascati, Italy), where a part of this work was done.
V.V.V. was supported by De Sitter cosmology fellowship and by the grant No.
18T-1C355 of the Committee of Science of the Ministry of Education and
Science RA.

\appendix

\section{Summation formula over the zeros of combinations of cylinder
functions}

\label{sec:App}

In this appendix we derive a summation formula over the positive zeros $%
x=\lambda _{\mu ,n}^{(\delta )}$, $n=1,2,\ldots $, of the function $g_{\mu
+\delta ,\mu }(x,\eta x)$, with $\delta =0,\pm 1$ and $\eta >1$, by using
the more general result from \cite{Saha87,Saha08}. Note that the equation $%
g_{\mu +\delta ,\mu }(x,\eta x)=0$ includes the equations for the
eigenvalues of the radial quantum number $\lambda $ for the boundary
conditions on a fermionic field we have discussed above. Namely, one should
take $\mu =ma-1/2$, $\delta =1$ for the condition (\ref{Bagbc}), $\mu
=ma+1/2 $, $\delta =-1$ for the condition (\ref{BC2}), and $\mu =ma\pm 1/2$,
$\delta =0$ for the remaining two boundary conditions discussed in Section %
\ref{sec:Brane}. In \cite{Saha87,Saha08}, on the base of the generalized
Abel-Plana formula, a summation formula is derived for the series over zeros
of the function $\bar{J}_{\mu }^{(a)}(x)\bar{Y}_{\mu }^{(b)}(\eta x)-\bar{Y}%
_{\mu }^{(a)}(x)\bar{J}_{\mu }^{(b)}(\eta x)$ with the notations $\bar{f}%
_{\mu }^{(j)}(z)=A_{j}f(z)+B_{j}zf^{\prime }(z)$, where $j=a,b$, and $A_{j}$%
, $B_{j}$ are constants. We take in that formula special values $A_{b}=1$, $%
B_{b}=0$, $A_{a}=\mu ^{|\delta |}$, $B_{a}=-\delta $. By using the
recurrence relations for the modified Bessel functions the following formula
is obtained%
\begin{eqnarray}
&&\sum_{n=1}^{\infty }h(\lambda _{\mu ,n}^{(\delta )})T_{\mu }^{(\delta
)}(\eta ,\lambda _{\mu ,n}^{(\delta )})=\frac{2}{\pi ^{2}}\int_{0}^{\infty }%
\frac{h(x)dx}{J_{\mu +\delta }^{2}(x)+Y_{\mu +\delta }^{2}(x)}  \notag \\
&&\qquad +\frac{1}{2\pi }\int_{0}^{\infty }dx\,\frac{\left[ h(xe^{\pi
i/2})+h(xe^{-\pi i/2})\right] K_{\mu }(\eta x)/K_{\mu +\delta }(x)}{I_{\mu
+\delta }(x)K_{\mu }(\eta x)-(-1)^{\delta }K_{\mu +\delta }(x)I_{\mu }(\eta
x)},  \label{SumFormAP}
\end{eqnarray}%
where $h(z)$ is an analytic function in the right half-plane of the complex
variable $z$, $I_{\mu }(x)$, $K_{\mu }(x)$ are the modified Bessel functions
and%
\begin{equation}
T_{\mu }^{(\delta )}(\eta ,x)=\frac{x}{J_{\mu +\delta }^{2}(x)/J_{\mu
}^{2}(\eta x)-1}.  \label{Tmud}
\end{equation}%
Note that the function in the denominator of the second integral in (\ref%
{SumFormAP}) is equal to $G_{\mu +\delta ,\mu }(x,\eta x)$ (see (\ref{Gnumu}%
)).

The function $h(z)$ may have branch points on the imaginary axis that should
be avoided by small semicircles in the right half-plane. Depending on the
behavior of the function $h(z)$ near the origin, a residue term at $z=0$ may
be present in the right-hand side of (\ref{SumFormAP}) (see \cite{Saha08}).
The corresponding contribution to the current density is cancelled by the
contribution of the fermionic zero mode (for the case of a scalar field see
\cite{Bell16}). By using the relation between the functions $I_{\pm \mu }(x)$
and $K_{\mu }(x)$, it can bee seen that for $\delta =0$ one gets $G_{\mu
,\mu }(x,\eta x)=G_{|\mu |,|\mu |}(x,\eta x)<0$. For $\delta =\pm 1$ and $%
\mu \geqslant 0$ one has $G_{\mu +\delta ,\mu }(x,\eta x)>0$. In particular,
from here it follows that for the boundary conditions we have discussed
above and for $ma\geqslant 0$ there are no fermionic modes with purely
imaginary $\lambda $.


\begin{thebibliography}{99}
\bibitem{Birr82} N.D. Birrell and P. C.W. Davies, \textit{Quantum Fields in
Curved Space} (Cambridge University Press, Cambridge, England, 1982); A.A.
Grib, S.G. Mamayev, and V. M. Mostepanenko, \textit{Vacuum Quantum Effects
in Strong Fields} (Friedmann Laboratory Publishing, St. Petersburg, 1994);
L.E. Parker and D.J. Toms, \textit{Quantum Field Theory in Curved Spacetime}
(Cambridge University Press, Cambridge, England, 2009).

\bibitem{Most97} V.M.~Mostepanenko and N.N.~Trunov, \textit{The Casimir
Effect and Its Applications} (Oxford University Press, Oxford, 1997);
K.A.~Milton, \textit{The Casimir Effect: Physical Manifestation of
Zero-Point Energy} (World Scientific, Singapore, 2002); M. Bordag, G.L.
Klimchitskaya, U. Mohideen, and V.M. Mostepanenko, \textit{Advances in the
Casimir Effect} (Oxford University Press, Oxford, 2009); \textit{Casimir
Physics}, edited by D. Dalvit, P. Milonni, D. Roberts, and F. da Rosa,
Lecture Notes in Physics Vol. 834 (Springer-Verlag, Berlin, 2011).

\bibitem{Maar10} R.~Maartens and K. Koyama, Living Rev. Relativity \textbf{13%
}, 5 (2010).

\bibitem{Ahar00} O.~Aharony, S.S.~Gubser, J.~Maldacena, H.~Ooguri, and
Y.~Oz, Phys. Rep. \textbf{323}, 183 (2000); H. N\u{a}stase, \textit{%
Introduction to AdS/CFT correspondence} (Cambridge University Press,
Cambridge, 2015); M. Ammon and J. Erdmenger, \textit{Gauge/Gravity Duality:
Foundations and Applications} (Cambridge University Press, Cambridge, 2015).

\bibitem{Pire14} A.S.T. Pires, AdS/CFT Correspondence in Condensed Matter
(Morgan \& Claypool Publishers, USA, 2014); J. Zaanen, Y.-W. Sun, Y. Liu,
and K. Schalm, Holographic Duality in Condensed Matter Physics (Cambridge
University Press, Cambridge, 2015); R.-G. Cai, L. Li, L.-F. Li, and R.-Q.
Yang, Sci. China-Phys. Mech. Astron. \textbf{58}(6), 060401 (2015); E.
Kiritsis and L. Li, J. High Energy Phys. \textbf{01} (2016) 147.

\bibitem{Fabi00} M. Fabinger and P. Horava, Nucl. Phys. B \textbf{580}, 243
(2000); S. Nojiri, S. Odintsov, and S. Zerbini, Phys. Rev. D 62, 064006
(2000); S. Nojiri, O. Obregon, and S. Odintsov, Phys. Rev. D \textbf{62},
104003 (2000); D. J. Toms, Phys. Lett. B \textbf{484}, 149 (2000); W.D.
Goldberger and I.Z. Rothstein, Phys. Lett. B \textbf{491}, 339 (2000); S.
Nojiri and S. Odintsov, J. High Energy Phys. 07 (2000) 049; J. Garriga, O.
Pujol\`{a}s, and T. Tanaka, Nucl. Phys. B \textbf{605}, 192 (2001); I.H.
Brevik, K.A. Milton, S. Nojiri, and S.D. Odintsov, Nucl. Phys. B \textbf{599}%
, 305 (2001); A. Flachi and D.J. Toms, Nucl. Phys. B \textbf{610}, 144
(2001); A.A. Saharian and M.R. Setare, Phys. Lett. B \textbf{552}, 119
(2003). E. Elizalde, S. Nojiri, S. D. Odintsov, and S. Ogushi, Phys. Rev. D
\textbf{67}, 063515 (2003); J. Garriga and A. Pomarol, Phys. Lett. B \textbf{%
560}, 91 (2003); R. A. Knapman and D. J. Toms, Phys. Rev. D \textbf{69},
044023 (2004); A. A. Saharian, Nucl. Phys. B \textbf{712}, 196 (2005); R.
Durrer and M. Ruser, Phys. Rev. Lett. \textbf{99}, 071601 (2007); M. Ruser
and R. Durrer, Phys. Rev. D \textbf{76}, 104014 (2007); A.A. Saharian and
A.L. Mkhitaryan, 08, 063 (2007); M. Frank, I. Turan and L. Ziegler, Phys.
Rev. D \textbf{76}, 015008 (2007); A. Flachi and T. Tanaka, Phys. Rev. D
\textbf{80}, 124022 (2009); L.P. Teo, Phys. Lett. B \textbf{682}, 259
(2009); M. Rypestol and I. Brevik, New. J. Phys. \textbf{12}, 013022 (2010);
R. Obousy and G. Cleaver, J. Geom. Phys. \textbf{61}, 577 (2011); E.R.
Bezerra de Mello, A.A. Saharian, and M.R. Setare, Phys. Rev. D \textbf{92},
104005 (2015); N. Haba and T. Yamada, arXiv: 1903.10160.

\bibitem{Flac01} A. Flachi, I.G. Moss, and D.J. Toms, Phys. Lett. B \textbf{%
518}, 153 (2001).

\bibitem{Flac01b} A. Flachi, I.G. Moss, and D.J. Toms, Phys. Rev. D \textbf{%
64}, 105029 (2001).

\bibitem{Shao10} S.-H. Shao, P. Chen and J.-A. Gu, Phys. Rev. D \textbf{81},
084036 (2010).

\bibitem{Eliz13} E. Elizalde, S.D. Odintsov, and A.A. Saharian, Phys. Rev. D
\textbf{87}, 084003 (2013); L.P. Teo, Int. J. Mod. Phys. A \textbf{28},
1350158 (2013).

\bibitem{Teo10} L.P. Teo, J. High Energy Phys. 10 (2010) 019; A.S. Kotanjyan
and A.A. Saharian, Phys. Atom. Nucl. \textbf{80}, 562 (2017); A.S.
Kotanjyan, A.A. Saharian, H.G. Sargsyan, and D.H. Simonyan, Proceedings of
Science PoS(MPCS2015)021.

\bibitem{Noji00} S. Nojiri and S. Odintsov, Phys. Lett. B \textbf{484}, 119
(2000); W. Naylor and M. Sasaki, Phys. Lett. B \textbf{542}, 289 (2002); E.
Elizalde, S. Nojiri, S.D. Odintsov, and S. Ogushi, Phys. Rev. D \textbf{67},
063515 (2003); I.G. Moss, W. Naylor, W. Santiago-Germ\'{a}n, and M. Sasaki,
Phys. Rev. D \textbf{67}, 125010 (2003); O. Pujol\`{a}s and T. Tanaka, J.
Cosmol. Astropart. Phys. 12 (2004) 009; A. Flachi, A. Knapman, W. Naylor,
and M. Sasaki, Phys. Rev. D \textbf{70}, 124011 (2004); W. Naylor and M.
Sasaki, Prog. Theor. Phys. \textbf{113}, 535 (2005); O. Pujol\`{a}s and M.
Sasaki, J. Cosmol. Astropart. Phys. 09 (2005) 002.

\bibitem{Flac03} A. Flachi, J. Garriga, O. Pujol\`{a}s, and T. Tanaka, J.
High Energy Phys. \textbf{08} (2003) 053; A. Flachi and O. Pujol\`{a}s,
Phys. Rev. D \textbf{68}, 025023 (2003); A.A. Saharian, Phys. Rev. D \textbf{%
73}, 044012 (2006); A.A. Saharian, Phys. Rev. D \textbf{73}, 064019 (2006);
A.A. Saharian, Phys. Rev. D \textbf{74}, 124009 (2006); E. Elizalde, M.
Minamitsuji, and W. Naylor, Phys. Rev. D \textbf{75}, 064032 (2007); R.
Linares, H.A. Morales-T\'{e}cotl, and O. Pedraza, Phys. Rev. D \textbf{77},
066012 (2008); M. Frank, N. Saad, and I. Turan, Phys. Rev. D \textbf{78},
055014 (2008).

\bibitem{Saha04} A.A. Saharian, Phys. Rev. D \textbf{70}, 064026 (2004);
A.A. Saharian, Phys. Rev. D \textbf{74}, 124009 (2006); A.A. Saharian and
H.G. Sargsyan, Astrophysics \textbf{61}, 375 (2018).

\bibitem{Beze13c} E.R. Bezerra de Mello and A.A. Saharian, Phys. Rev. D
\textbf{87}, 045015 (2013).

\bibitem{Bell10} S. Bellucci, A.A. Saharian, and V.M. Bardeghyan, Phys. Rev.
D \textbf{82}, 065011 (2010).

\bibitem{Bell14} S. Bellucci, E.R. Bezerra de Mello, and A.A. Saharian,
Phys. Rev. D \textbf{89}, 085002 (2014).

\bibitem{Bell13} S. Bellucci and A.A. Saharian, Phys. Rev. D \textbf{87},
025005 (2013).

\bibitem{Bell15} S. Bellucci, A.A. Saharian, and N.A. Saharyan, Eur. Phys.
J. C \textbf{75}, 378 (2015).

\bibitem{Rech07} P. Recher, B. Trauzettel, A. Rycerz, Y.M. Blanter, C.W.J.
Beenakker, and A.F. Morpurgo, Phys. Rev. B \textbf{76}, 235404 (2007); S.
Bellucci, A.A. Saharian, and A.Kh. Grigoryan, Phys. Rev. D \textbf{94},
105007 (2016).

\bibitem{Bluh09} H. Bluhm, N. Koshnick, J. Bert, M. Huber, and K. Moler,
Phys. Rev. Lett. \textbf{102}, 136802 (2009); A.C. Bleszynski-Jayich, W. E.
Shanks, B. Peaudecerf, E. Ginossar, F. von Oppen, L. Glazman, and J.G.E.
Harris, Science \textbf{326}, 272 (2009).

\bibitem{Beze10C} E.R. Bezerra de Mello, V. Bezerra, A.A. Saharian, and V.M.
Bardeghyan, Phys. Rev. D \textbf{82}, 085033 (2010); S. Bellucci, E.R.
Bezerra de Mello, and A. A. Saharian, Phys. Rev. D \textbf{83}, 085017
(2011); A.A. Saharian, E.R. Bezerra de Mello, and A.A. Saharyan, arXiv:
1907.04196.

\bibitem{Bell13b} S. Bellucci, A.A. Saharian, and H.A. Nersisyan, Phys. Rev.
D \textbf{88}, 024028 (2013).

\bibitem{Beze15} E.R. Bezerra de Mello and A.A. Saharian, and V. Vardanyan,
Phys. Lett. B \textbf{741}, 155 (2015).

\bibitem{Bell17} S. Bellucci, A.A. Saharian, and V. Vardanyan, Phys. Rev. D
\textbf{96}, 065025 (2017).

\bibitem{Oliv18} W. Oliveira dos Santos, H. F. Mota, and E. R. Bezerra de
Mello, Phys. Rev. D \textbf{99}, 045005 (2019).

\bibitem{Bell15b} S. Bellucci, A.A. Saharian, and V. Vardanyan, J. High
Energy Phys. 11 (2015) 092.

\bibitem{Bell16} S. Bellucci, A.A. Saharian, and V. Vardanyan, Phys. Rev. D
\textbf{93}, 084011 (2016).

\bibitem{Bell18} S. Bellucci, A.A. Saharian, D.H. Simonyan, and V.
Vardanyan, Phys. Rev. D \textbf{98}, 085020 (2018).

\bibitem{Prud86} A.P.~Prudnikov, Yu.A.~Brychkov, and O.I.~Marichev, \textit{%
Integrals and series} (Gordon and Breach, New York, 1986), Vol.2.

\bibitem{Ambr15} V.E. Ambru\c{s} and E. Winstanley, Phys. Lett. B \textbf{749%
}, 597 (2015); V.E. Ambru\c{s} and E. Winstanley, Class. Quantum Grav.
\textbf{34}, 145010 (2017).

\bibitem{Abra72} \textit{Handbook of Mathematical Functions}, edited by M.
Abramowitz and I.A. Stegun (Dover, New York, 1972).

\bibitem{Berr87} M.V. Berry and R.J. Mondragon, Proc. R. Soc. A \textbf{412}%
, 53 (1987).

\bibitem{Rand99} L. Randall and R. Sundrum, Phys. Rev. Lett. \textbf{83},
3370 (1999); L. Randall and R. Sundrum, Phys. Rev. Lett. \textbf{83}, 4690
(1999).

\bibitem{Chan05} S. Chan, S.Ch. Park, and J. Song, Phys. Rev. D \textbf{71},
106004 (2005).

\bibitem{Gusy07} V.P. Gusynin, S.G. Sharapov, and J.P. Carbotte, Int. J.
Mod. Phys. B \textbf{21}, 4611 (2007); A.H. Castro Neto, F. Guinea, N.M.R.
Peres, K.S. Novoselov, and A.K. Geim, Rev. Mod. Phys. \textbf{81}, 109
(2009).

\bibitem{Gupt18} S. Gupta and A. Saxena (Editors), \textit{The Role of
Topology in Materials} (Springer, Switzerland, 2018).

\bibitem{Kole09} D.V. Kolesnikov and V. A. Osipov, Phys. Part. Nucl. \textbf{%
40}, 502 (2009); M.A.H. Vozmediano, M.I. Katsnelson, and F. Guinea, Phys.
Rept. \textbf{496}, 109 (2010).

\bibitem{Iori14} A. Iorio and G. Lambiase, Phys. Rev. D \textbf{90}, 025006
(2014).

\bibitem{Morr19} T. Morresi, D. Binosi, S. Simonucci, R. Piergallini, S.
Roche, N.M. Pugno, and S. Taioli, arXiv:1907.08960.

\bibitem{Gonz10} J. Gonz\'{a}lez and J. Herrero, Nucl. Phys. B \textbf{825},
426 (2010); V. Atanasov and A. Saxena, J. Phys.: Condens. Matter \textbf{23}%
, 175301 (2011); R. Pincak and J. Smotlacha, Eur. Phys. J. B \textbf{86},
480 (2013); A. Sepehri, R. Pincak, K. Bamba, S. Capozziello, and E.N.
Saridakis, Int. J. Mod. Phys. D \textbf{26}, 1750094 (2017).

\bibitem{Volo15} G.E. Volovik and M.A. Zubkov, Ann. Phys. \textbf{356}, 255
(2015); B. Amorim et. al., Phys. Rept. \textbf{617}, 1 (2016).

\bibitem{Inag97} T. Inagaki, T. Muta, and S.D. Odintsov, Prog. Theor. Phys.
Suppl. 127, 93 (1997); A. Flachi, Phys. Rev. D \textbf{86}, 104047 (2012);
A. Flachi, Phys. Rev. D \textbf{88}, 085011 (2013); A. Flachi, M. Nitta, S.
Takada, and R. Yoshii, Phys. Rev. Lett. \textbf{119}, 031601 (2017); A.
Flachi and V. Vitagliano, Phys. Rev. D \textbf{99}, 125010 (2019).

\bibitem{Cort12} A. Cortijo, F. Guinea, and M.A.H. Vozmediano, J. Phys. A
\textbf{45}, 383001 (2012); V.N. Kotov, B. Uchoa, V.M. Pereira, F. Guinea,
and A.H. Castro Neto, Rev. Mod. Phys. \textbf{84},1067 (2012).

\bibitem{Saha87} A.A. Saharian, Izvestiia Akademii nauk Armianskoi SSR
Matematika \textbf{22}, 166 (1987) [English translation: A.A. Saaryan, Sov.
J. Contemp. Math. Anal. \textbf{22}, 70 (1987)].

\bibitem{Saha08} A.A. Saharian, \textit{The Generalized Abel-Plana Formula
with Applications to Bessel Functions and Casimir Effect} (Yerevan State
University Publishing House, Yerevan, 2008); Report No. ICTP/2007/082;
arXiv:0708.1187.
\end{thebibliography}
\end{document}